# Disequilibrium biosignatures over Earth history and implications for detecting exoplanet life

Joshua Krissansen-Totton,[1,2]* Stephanie Olson,[3] David C. Catling[1,2]

Chemical disequilibrium in planetary atmospheres has been proposed as a generalized method for detecting life on exoplanets through remote spectroscopy. Among solar system planets with substantial atmospheres, the modern Earth has the largest thermodynamic chemical disequilibrium due to the presence of life. However, how this disequilibrium changed over time and, in particular, the biogenic disequilibria maintained in the anoxic Archean or less oxic Proterozoic eons are unknown. We calculate the atmosphere-ocean disequilibrium in the Precambrian using conservative proxy- and model-based estimates of early atmospheric and oceanic compositions. We omit crustal solids because subsurface composition is not detectable on exoplanets, unlike above-surface volatiles. We find that (i) disequilibrium increased through time in step with the rise of oxygen; (ii) both the Proterozoic and Phanerozoic may have had remotely detectable biogenic disequilibria due to the coexistence of $O_2$, $N_2$, and liquid water; and (iii) the Archean had a biogenic disequilibrium caused by the coexistence of $N_2$, $CH_4$, $CO_2$, and liquid water, which, for an exoplanet twin, may be remotely detectable. On the basis of this disequilibrium, we argue that the simultaneous detection of abundant $CH_4$ and $CO_2$ in a habitable exoplanet's atmosphere is a potential biosignature. Specifically, we show that methane mixing ratios greater than $10^{-3}$ are potentially biogenic, whereas those exceeding $10^{-2}$ are likely biogenic due to the difficulty in maintaining large abiotic methane fluxes to support high methane levels in anoxic atmospheres. Biogenicity would be strengthened by the absence of abundant CO, which should not coexist in a biological scenario.

## INTRODUCTION

Life produces waste gases that modify an atmosphere's composition, and it will soon be possible to look for such biosignature gases on exoplanets using telescopic observations. In the near future, a high-contrast imaging system coupled to a spectrograph on the Very Large Telescope may allow the detection of biosignature gases on the nearest exoplanets (1). The James Webb Space Telescope (JWST), scheduled to launch in 2019, will search for biosignature gases on transiting exoplanets such as the TRAPPIST-1 system (2, 3). In the 2020s, three large ground telescopes—the European Extremely Large Telescope, the Thirty Meter Telescope, and the Giant Magellan Telescope—could look for oxygen, water vapor, and carbon dioxide on nearby exoplanets (4–6), whereas the Wide-Field Infrared Survey Telescope may also be capable of detecting biosignature gases on planets orbiting close stars (7). In the more distant future, next-generation space telescopes could survey our stellar neighborhood for biosignatures (8, 9).

Considerable attention has been given to $O_2$ as a biosignature gas because it is challenging to produce in large quantities without oxygenic photosynthesis (10). Although several pathological scenarios have been proposed whereby a terrestrial planet in the habitable zone might accumulate abiotic oxygen, additional contextual information can rule out these false-positive scenarios (10), whereas some doubt the realism of some of these false-positive concepts (11).

However, even if oxygenic photosynthesis is present, it does not guarantee detectable levels of atmospheric oxygen. There was virtually no oxygen in the Archean eon [4.0 to 2.5 billion years ago (Ga)] (12) despite the possible origin of oxygenic photosynthesis by 3.0 Ga (13).

Oxygen levels in the Proterozoic eon (2.5 to 0.541 Ga) are disputed, but some proxy estimates imply remotely undetectable levels (14). More generally, we do not know whether oxygenic photosynthesis is a likely evolutionary development. Oxygenic photosynthesis is enzymatically complex and only evolved once on Earth (15).

For all these reasons, alternative approaches to biosignatures are needed. Previous studies have explored hydrocarbons and their hazes, organosulfur compounds, and biological pigments as biosignatures for anoxic worlds similar to the early Earth [reviewed by Schwieterman et al. (16)], but these approaches rely on specific metabolisms with high productivities.

A more general biosignature, which has not been considered for the early Earth, is atmospheric chemical disequilibrium, evident in the coexistence of two or more long-term incompatible gases (17–20). The modern $O_2$-$CH_4$ redox couple is widely believed to be a compelling disequilibrium biosignature because of the short kinetic lifetime (~10 years) of methane in Earth's atmosphere (21, 22), which requires a substantial source flux of $CH_4$ in excess of reasonable abiogenic sources.

A number of arguments against the concept of disequilibrium biosignatures have been proposed. Kleidon [(23), p. 250] notes that atmospheric disequilibrium between $O_2$ and $CH_4$ exists because of incomplete decomposition of organic matter. The power associated with this leakage of unused free energy is a small fraction (0.3%) of power involved in photosynthesis, and so, it is argued that atmospheric chemical disequilibrium is not a good indicator of biospheric activity. However, chemical disequilibrium need not map to the amount of biological production to be a good biosignature—it merely has to reveal the existence of life. The methane flux required to sustain observed quantities of methane in the modern Earth's oxidizing atmosphere is greater than what abiotic processes could plausibly provide, and thus, biological methane leakage must be invoked to explain the persistent disequilibrium. In addition, it has been argued that the disequilibrium in the Earth's atmosphere is merely a reflection of high oxygen levels and that statements about disequilibrium therefore reduce to statements about oxygen (24). However, in this paper, we show that there was an important biogenic disequilibrium in the anoxic Archean atmosphere.

[1]Department of Earth and Space Sciences/Astrobiology Program, University of Washington, Seattle, WA 98195, USA. [2]Virtual Planetary Laboratory, University of Washington, Seattle, WA 98195, USA. [3]Department of Earth Sciences and NASA Astrobiology Institute, University of California, Riverside, Riverside, CA 92521, USA.
*Corresponding author. Email: joshkt@uw.edu

Previously, we quantified the thermodynamic disequilibrium in solar system atmospheres by taking observed compositions, reacting them numerically to chemical thermodynamic equilibrium, and calculating the Gibbs free energy difference between observed and equilibrium states (25). Earth's purely gas-phase disequilibrium, as measured by available Gibbs free energy, is 1.5 J/mol of atmosphere and largely attributable to the $O_2$-$CH_4$ thermodynamic disequilibrium. The magnitude of this disequilibrium is not large compared to the abiogenic disequilibrium calculated for other solar system bodies, but given the kinetic considerations discussed above, $CH_4$ and $O_2$ remain a robust disequilibrium biosignature. In contrast, when the Earth's entire fluid envelope (the atmosphere-ocean reservoir) is considered, the disequilibrium is very large (2326 J/mol) due to the coexistence of $O_2$, $N_2$, and liquid water. These three species should react to form nitric acid in thermodynamic equilibrium. Krissansen-Totton et al. (25) discuss how this biogenic disequilibrium is potentially detectable on exoplanets similar to the modern Earth.

Although Earth's chemical disequilibrium is large today, how the disequilibrium changed through Earth history has not been quantified—whether it was large in the Precambrian when there was less atmospheric oxygen, and whether biogenic species would have contributed in an anoxic atmosphere. The $N_2$-$O_2$-$H_2O$ and $O_2$-$CH_4$ disequilibrium biosignatures may not have been present for much of Earth history because the atmosphere had 20 ± 10% $O_2$ for only the last one-eighth of its history and virtually no $O_2$ in the Archean [reviewed by Catling and Kasting (26), chap. 10]. Consequently, we seek to calculate the thermodynamic disequilibrium for the Archean, Proterozoic, and Phanerozoic atmosphere-ocean systems. We explore how Earth's atmosphere-ocean disequilibrium has coevolved with life and find that the results suggest a novel disequilibrium biosignature for Archean-like exoplanets.

## RESULTS
### Approach for calculating thermodynamic disequilibrium

We calculated chemical thermodynamic disequilibrium in the atmosphere-ocean system according to the methodology shown schematically in Fig. 1 and fully described in Materials and Methods. Our MATLAB code is available on the website of the lead author. Given an assumed composition for the atmosphere and ocean of the early Earth, we react the whole system to thermodynamic equilibrium using Gibbs energy minimization. The equilibrium abundances of reactive constituents differ from the initial abundances, but atoms and charge are conserved. We neglect solids and most nonvolatile aqueous species because our focus is on remotely observable disequilibria (justified further in Discussion and in section S1).

To quantify the chemical thermodynamic disequilibrium in a planet's atmosphere-ocean system, we define the "available Gibbs energy" as the difference in Gibbs free energy between the initial (observed) state and the equilibrium state

$$\Phi \equiv G_{(T,P)}(\mathbf{n}_{\text{initial}}) - G_{(T,P)}(\mathbf{n}_{\text{final}}) \quad (1)$$

The available Gibbs energy, $\Phi$, has units of joules per mole of atmosphere. The vector $\mathbf{n}_{\text{initial}}$ contains the abundances of all the atmospheric and ocean constituents of the initial state, whereas $\mathbf{n}_{\text{final}}$ contains abundances of the final state. This Gibbs free energy difference is the maximum useful work that can be extracted from the system. That is, $\Phi$ is the untapped chemical free energy in a planet's atmosphere and so provides our metric of disequilibrium. Note that when we discuss life exploiting the free energy in a planet's atmosphere, we are referring to surface (or subsurface) life consuming atmospheric gases. Although there are microbes that are adapted to survival in the upper troposphere (27), no known organism subsists independently of the surface.

Table 1 shows estimates of the composition of the atmosphere and ocean in the Precambrian, adopted as initial abundances in our

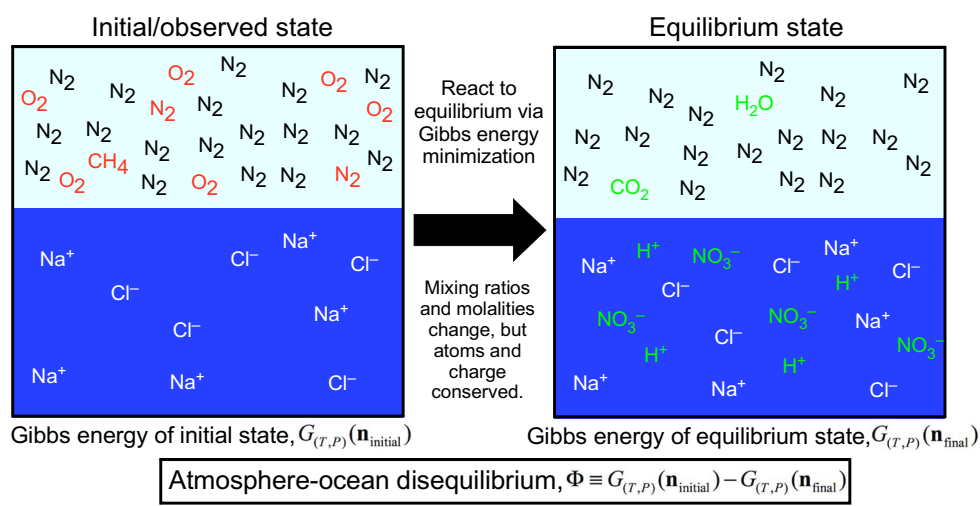

**Fig. 1. Schematic of methodology for calculating atmosphere-ocean disequilibrium.** We quantify the disequilibrium of the atmosphere-ocean system by calculating the difference in Gibbs energy between the initial and final states. The species in this particular example show the important reactions to produce equilibrium for the Phanerozoic atmosphere-ocean system, namely, the reaction of $N_2$, $O_2$, and liquid water to form nitric acid, and methane oxidation to $CO_2$ and $H_2O$. Red species denote gases that change when reacted to equilibrium, whereas green species are created by equilibration. Details of aqueous carbonate system speciation are not shown.

Table 1. **Assumed initial atmosphere-ocean composition for Archean and Proterozoic.**

| Atmospheric species | Archean range | | | Proterozoic range | | |
|---|---|---|---|---|---|---|
| | Mixing ratio | | Reference/explanation | Mixing ratio | | Reference/explanation |
| | Minimum disequilibrium | Maximum disequilibrium | | Minimum disequilibrium | Maximum disequilibrium | |
| $N_2(g)$ | 0.98 | 0.5 | Mixing ratios sum to 1 | 0.99 | 0.86 | Mixing ratios sum to 1 |
| $O_2(g)$ | $1 \times 10^{-10}$ | $2 \times 10^{-7}$ | (39) | 0.0001 | 0.03 | (14, 76) |
| $CH_4(g)$ | 0.0001 | 0.01 | (77) | $3 \times 10^{-6}$ | $1 \times 10^{-4}$ | (78, 79) |
| $CO_2(g)$ | 0.001 | 0.5 | (80, 81) | 0.0001 | 0.1 | (80, 81) |
| $H_2(g)$ | 0 | 0.0001 | (59) | 0 | $2 \times 10^{-6}$ | (82) |
| $N_2O(g)$ | 0 | 0 | No denitrification so negligible production | 0 | $1 \times 10^{-6}$ | (79, 83) |
| $NH_3(g)$ | 0 | $1 \times 10^{-9}$ | (84) | 0 | 0 | Negligible in bulk atmosphere |
| $O_3(g)$ | 0 | 0 | Negligible in bulk atmosphere | 0 | 0 | Negligible in bulk atmosphere (79) |
| $CO(g)$ | 0 | 0.001 | (59) | 0 | $2 \times 10^{-7}$ | (82) |
| **Ocean species** | Molality (mmol/kg) | | Reference/explanation | Molality (mmol/kg) | | Reference/explanation |
| | Minimum disequilibrium | Maximum disequilibrium | | Minimum disequilibrium | Maximum disequilibrium | |
| $Na^+$ | 550 | 586 | Charge balance | 547 | 549 | Charge balance |
| $Cl^-$ | 546 | 546 | Modern value | 546 | 546 | Modern value |
| $SO_4^{2-}$ | 0 | 0.2 | (85, 86) | 0.25 | 5 | (87, 88) |
| $H_2S$ | 0 | 0.004 | In euxinic oceans using a Black Sea analog* | 0 | 0.004 | In euxinic oceans using a Black Sea analog* |
| $NH_4^+$ | 0 | 0.050 | Set by phosphorus† assuming Redfield ratios and the presence of N fixation, given that N fixation evolved early (89) | 0 | 0.050 | Set by phosphorus† assuming Redfield ratios |
| $NO_3^-$ | 0 | 0 | Anoxic bulk ocean | 0 | 0 | Anoxic bulk ocean |
| Alkalinity | 4 | 40 | (29); Krissansen-Totton et al., in preparation | 1.0 | 3.0 | (29); Krissansen-Totton et al., in preparation |
| pH | 8.0 | 6.3 | Carbon chemistry equilibrium‡ | 8.4 | 6.0 | Carbon chemistry equilibrium‡ |

*The concentration of $H_2S$ in the Black Sea is around 400 μmol/kg (90). However, <1 to 10% of the Precambrian seafloor was euxinic (91), which implies <1% euxinia by volume because euxinic continental slopes are much shallower than the deep ocean. Thus, we assumed 4 μmol/kg as an upper limit for bulk ocean aqueous $H_2S$.    †The maximum subsurface concentration of $NH_4^+$ is ultimately controlled by flux of phosphate from continental weathering. Assuming a modern dissolved phosphate abundance of ~2 μmol/kg, this implies an ammonium abundance of 32 μmol/kg assuming a 1:16 Redfield ratio. Results are largely insensitive to initial abundances of $NH_4^+$ (Fig. 5).    ‡pH is calculated from alkalinity and $P_{CO_2}$ assuming chemical equilibrium.

calculations. Plausible ranges for atmospheric mixing ratios and aqueous species molalities are taken from the literature, which includes proxy estimates and theoretical modeling. For both the Archean and Proterozoic, we calculate two end-member cases, denoted maximum and minimum disequilibrium. The maximum disequilibrium case assumes the largest possible mixing ratios and molalities of reactive species from the literature, whereas the minimum disequilibrium case assumes the converse. In the assumed initial ocean chemistry, [$Na^+$], a conservative nonreactive ion, is adjusted to achieve charge balance. Similarly, the initial mixing ratio of $N_2$ is adjusted in every case to ensure that mixing ratios sum to unity.

Calculations were performed at 1-bar surface pressure for which partial pressures are numerically equivalent to mixing ratio constraints, but in the Supplementary Materials, we repeat our Archean calculations at both higher and lower pressures and find that changing atmospheric pressure has little effect on the results. Unless stated otherwise, initial dissolved gas abundances were calculated using Henry's law with coefficients from the National Institute of Standards and Technology (NIST) database (28). The errors introduced by assuming saturation for dissolved species are discussed in section S1 and found to be small. Initial water vapor abundances are also determined using Henry's law. In practice, tropospheric water vapor is spatially and temporally highly variable (0 to 4%) and controlled by the dynamics of the hydrological cycle, but using Henry's law yields an initial abundance (1.6%) consistent with this empirical range.

Ranges for ocean carbonate alkalinity, which is defined as the charge-weighted sum of carbon-bearing ions, $2[CO_3^{2-}] + [HCO_3^-]$, are based loosely on the study of Halevy and Bachan (29). However, in the Supplementary Materials, we investigate the sensitivity of our results to different assumed alkalinities and find that our key conclusions are unchanged.

Given alkalinity and atmospheric $P_{CO_2}$ (partial pressure of $CO_2$), we calculate ocean pH, carbonate, and bicarbonate concentrations from equilibrium chemistry. This procedure ensures that out-of-equilibrium carbon chemistry does not contribute to our disequilibrium calculations. Of course, in calculating disequilibria, carbonate speciation may be shifted by the reaction of other species in the system.

A commercial chemical engineering software package called Aspen Plus (version 8.6) was used to validate all the MATLAB calculations reported in this paper [see the study of Krissansen-Totton et al. (25) for full description of its implementation]. Tables comparing MATLAB and Aspen results are reported in section S2. In general, MATLAB and Aspen outputs agree to within 10% or better. Small differences are expected because the thermodynamic models in our MATLAB code differ from those in Aspen Plus. However, unlike the proprietary code, ours is open source and so fully transparent.

### Thermodynamic disequilibrium over Earth history

Here, we report results for our maximum and minimum disequilibrium in the Proterozoic and Archean. If the true atmosphere and ocean abundances are bounded by the values in Table 1, then the minimum and maximum disequilibria we calculate will encompass the true disequilibrium of the Earth's atmosphere-ocean system through time.

Figure 2 shows our calculated evolution of Earth's atmosphere-ocean disequilibrium. The modern atmosphere-ocean disequilibrium was analyzed at length by Krissansen-Totton et al. (25). In Fig. 2, the Phanerozoic Earth range was determined by using the abundances from (25) but varying initial oxygen mixing ratios from 0.1 to 0.3, which is the range inferred for the Phanerozoic (30). Two ranges are provided for the Proterozoic representing different assumptions about Proterozoic oxygen. We find that Earth's atmosphere-ocean disequilibrium was the smallest in the Archean, increased with the initial rise of oxygen during the Paleoproterozoic Great Oxidation Event, and then increased again after a second major increase in oxygen during the Neoproterozoic (Fig. 2).

The calculations that follow explain the evolution of Precambrian disequilibrium and which species are out of equilibrium and therefore contributing to the available Gibbs energy. This gives insight into how the disequilibria are affected by life and which species could serve as biosignatures.

### The Proterozoic disequilibrium and species that contribute to it

The available Gibbs energy for the maximum Proterozoic case is 884 J/mol, and the initial and equilibrium abundances for this case are shown in Fig. 3. We computed the contribution of individual reactions to this Gibbs energy by repeating the equilibrium calculation without the reaction products of specific reactions and by checking the results against semianalytic calculations (25).

The largest source of disequilibrium in the Proterozoic is the same as for the modern Earth: The levels of $N_2$, $O_2$, and liquid water should not coexist but rather react to form nitric acid

$$5O_2 + 2N_2 + 2H_2O \rightarrow 4H^+ + 4NO_3^- \quad (2)$$

The depletion of $O_2$ and the increase in $H^+$ and $NO_3^-$ are seen in Fig. 3. The formation of nitric acid also drives carbon speciation to a new equilibrium (section S3). Nitric acid formation and subsequent adjustment of carbon speciation contributes ~640 J/mol, the majority of Proterozoic atmosphere-ocean disequilibrium (72%).

Other reactions that contribute to the maximum Proterozoic disequilibrium are shown in section S3. Methane oxidation contributes considerably (75 J/mol) to the maximum Proterozoic disequilibrium because assumed methane abundances are much higher than on the modern Earth

$$CH_4 + 2O_2 \rightarrow 2H_2O + CO_2 \quad (3)$$

For the minimum Proterozoic case, the disequilibrium is still dominated by nitrate formation, but the available energy is only 9.5 J/mol due to lower initial $P_{O_2}$ (partial pressure of oxygen). Figure S1 shows the initial and equilibrium abundances for the minimum Proterozoic case,

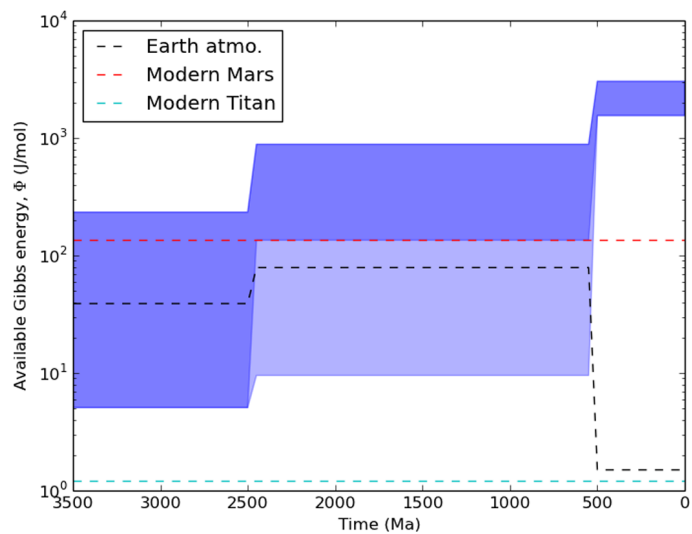

**Fig. 2. The evolution of Earth's atmosphere-ocean disequilibrium through time, as measured by available Gibbs free energy.** The blue shaded regions show the evolution of Earth's atmosphere-ocean disequilibrium. The wide ranges in the Archean and Proterozoic span our minimum and maximum disequilibrium scenarios. The large ranges are attributable to uncertainties in the atmospheric composition in each eon, mainly uncertain $P_{CH_4}$ in the Archean and uncertain $P_{O_2}$ in the Proterozoic. The two shadings for the Proterozoic represent different assumptions about atmospheric oxygen levels that represent divergent views in the current literature. Darker blue denotes $P_{O_2}$ > 2% PAL (present atmospheric level), whereas lighter blue denotes $P_{O_2}$ < 2% PAL. We calculate a secular increase in Earth's atmosphere-ocean disequilibrium over Earth history, correlated with the history of atmospheric oxygen. The black dashed line shows the upper bound of the Earth's atmosphere-only disequilibrium through time. We also include the modern (photochemically produced) disequilibria of Mars (red dashed) and Titan (blue dashed) for comparison (25). The abiotically produced disequilibria of all the other solar system planets are ≪1 J/mol (25).

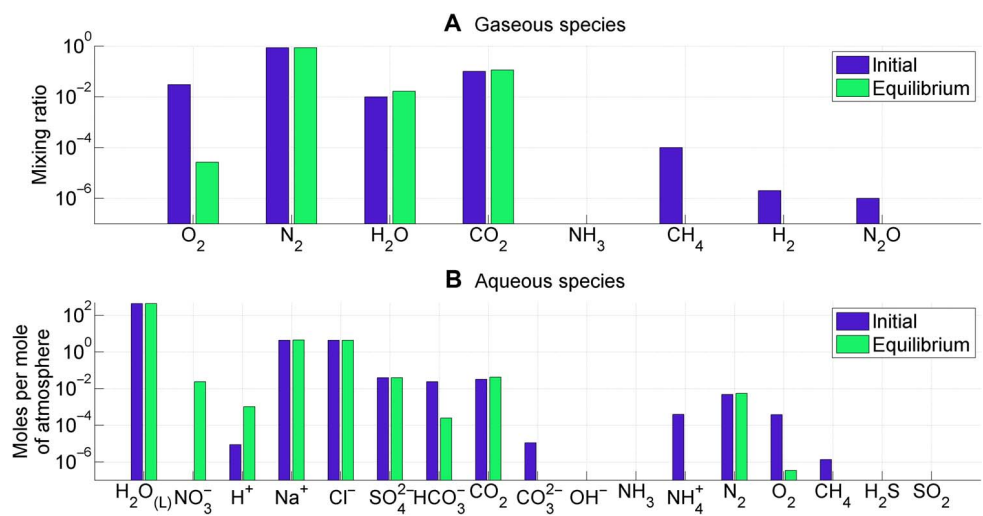

**Fig. 3. Atmosphere-ocean disequilibrium in the Proterozoic (maximum disequilibrium scenario).** Blue bars denote assumed initial abundances from the literature, and green bars denote equilibrium abundances calculated using Gibbs free energy minimization. Subplots separate (**A**) atmospheric species and (**B**) ocean species. The most important contribution to Proterozoic disequilibrium is the coexistence of atmospheric oxygen, nitrogen, and liquid water. These three species are lessened in abundance by reaction to equilibrium to form aqueous $H^+$ and $NO_3^-$. Changes in carbonate speciation caused by the decrease in ocean pH also contribute to the overall Gibbs energy change.

and tables S1 and S2 record numerical abundances for both maximum and minimum Proterozoic cases.

### The Archean disequilibrium and species that contribute to it

The initial and equilibrium abundances for the maximum Archean case are shown in Fig. 4. The available Gibbs energy for the maximum Archean case is 234 J/mol, which is dominated by the coexistence of $CO_2$, $N_2$, $CH_4$, and liquid water. These four species should not coexist but rather should react to form ammonium and bicarbonate, depleting almost all atmospheric methane (99.8% of the initial $CH_4$ is consumed by reaction to equilibrium)

$$5CO_2 + 4N_2 + 3CH_4 + 14H_2O \rightarrow 8NH_4^+ + 8HCO_3^- \quad (4)$$

The depletion of $CH_4$ and increase in $NH_4^+$ and $HCO_3^-$ are seen in Fig. 4. This reaction alone contributes ~170 J/mol (74%) of the maximum Archean disequilibrium. Other reactions that contribute to the maximum Archean disequilibrium are shown in section S3.

The available energy for the minimum Archean case is 5.1 J/mol, but even in this case, $CH_4$ is 99.99% depleted with respect to its initial abundance at equilibrium, demonstrating that $CH_4$ should not coexist in equilibrium with $N_2$-$CO_2$-$H_2O$ (liquid) across a broad range of initial conditions. Figure S2 shows the initial and equilibrium abundances for the minimum Archean case. Tables S4 and S5 also record numerical abundances for both maximum and minimum Archean cases.

### DISCUSSION
#### Disequilibria and the history of life

Our results show that Earth's disequilibrium has been strongly affected by life. The evolution of Earth's atmosphere-ocean disequilibrium follows the rise of biogenic oxygen (Fig. 2) similar to how the use of energy by life over Earth history was changed by an anaerobic to aerobic transition (*31*). Our calculated quantitative evolution of Earth's atmosphere-ocean thermodynamic disequilibrium is also consistent with qualitative speculations about the evolution of free energy dissipation by the biosphere through time [figure 12.5 of Kleidon (*23*)]. Oxygenic photosynthesis maintains disequilibrium in the Proterozoic and Phanerozoic by replenishing $O_2$ against $O_2$ sinks. Oxygenic photosynthesis also replenishes $N_2$ because the organic matter used in nitrate reduction and subsequent denitrification is produced by oxygenic photosynthesis (although approximately half of the $N_2$ replenishment comes from outgassing—see below for a more detailed discussion of nitrogen cycling). The emergence of oxygenic photosynthesis and the associated rise in primary productivity increased the disequilibrium of Earth's atmosphere-ocean system. Disequilibrium increased again in the Phanerozoic because oxygenic photosynthesis left a larger imprint on the environment following the Neoproterozoic rise of oxygen.

Before the advent of oxygenic photosynthesis, we calculate that Earth's disequilibrium was probably smaller than at any subsequent time (Fig. 2). If Archean life was exclusively chemotrophic, then it may have decreased a preexisting abiotic disequilibrium, mostly from the $H_2$-$CO_2$ pair in the atmosphere (*32*). With the advent of anoxygenic photosynthesis, the atmosphere-ocean disequilibrium may have increased because additional electron donors become available that were not limited by atmospheric abiotic disequilibrium. For example, Fe-oxidizing phototrophs produce organic carbon through the following net reaction

$$4Fe^{2+} + CO_2 + 11H_2O + h\nu \rightarrow 4Fe(OH)_3 + CH_2O + 8H^+ \quad (5)$$

The organic matter produced may then be converted to methane by anaerobic processing, thereby adding to the Archean atmosphere-ocean disequilibrium without the need for outgassed electron donors like $H_2$; methane is eventually photochemically oxidized to $CO_2$, thereby closing the cycle. We do not attempt to capture this change in Fig. 2 because we lack sufficient constraints about the advent of types of anoxygenic photosynthesis and their relative influence on atmospheric composition.

However, in both scenarios of anoxygenic photosynthesis and chemotrophy, the Archean biosphere does not drive the atmosphere-ocean system toward equilibrium. Instead, a $CH_4$-$N_2$-$H_2O$-$CO_2$ disequilibrium persists, maintained by methanogens (albeit smaller in the chemotrophic case). After the emergence of oxygenic photosynthesis, but before the rise of oxygen, the $CH_4$-$N_2$-$H_2O$-$CO_2$ disequilibrium may have increased further due to a larger biological $CH_4$ flux from the anaerobic processing of organic matter as the productivity of the biosphere increased.

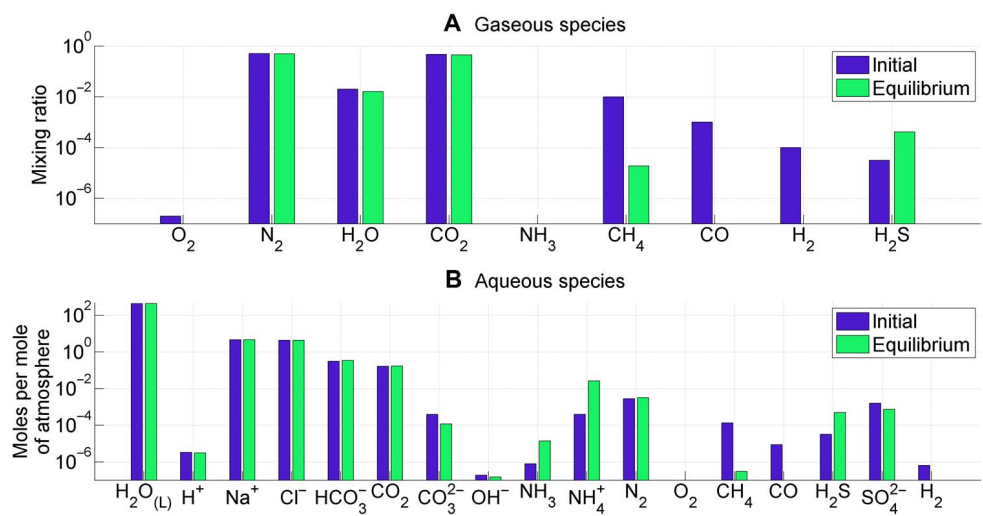

**Fig. 4. Atmosphere-ocean disequilibrium in the Archean (maximum disequilibrium scenario).** Blue bars denote assumed initial abundances from the literature, and green bars denote equilibrium abundances calculated using Gibbs free energy minimization. Subplots separate (**A**) atmospheric species and (**B**) ocean species. The most important contribution to Archean disequilibrium is the coexistence of atmospheric $CH_4$, $N_2$, $CO_2$, and liquid water. These four species are lessened in abundance by reaction to equilibrium to form aqueous $HCO_3^-$ and $NH_4^+$. Oxidation of CO and $H_2$ also contributes to the overall Gibbs energy change.

Figure 2 also shows the evolution of Earth's atmosphere-only disequilibrium (bounded by the black dashed line). The atmosphere-only disequilibrium decreases with time and thus does not reflect the rise of oxygen or the growth of primary productivity since 3.5 Ga. This illustrates the importance of including the oceans when quantifying disequilibrium in fluid reservoirs.

An interesting question emerges from this analysis: Given that Earth's atmosphere-ocean system has been massively out of equilibrium since the early Archean, why has life not evolved to consume this "free lunch"? In particular, why have no metabolisms evolved to exploit the $N_2$-$O_2$-$H_2O$ disequilibrium present after the Archean? Nitrogen-fixing bacteria and nitrifiers convert $N_2$ to nitrate, and although the nitrification step yields free energy, no known organism obtains energy by combining these two reactions. Lewis and Randall [(*33*), pp. 567–568] were the first to recognize that $N_2$, $O_2$, and water were out of equilibrium in the Earth system, and they argued that it was fortunate for us that life had not evolved to catalyze their reaction because, otherwise, the atmosphere would have been depleted of oxygen and the oceans turned to dilute nitric acid.

In reality, it is hard to predict what the end result of such evolutionary innovation would be. Atmospheric oxygen would be replenished to some extent by photosynthesis, but nitrogen might be drawn down faster than it could be replenished because there are abundant cations in the crust to neutralize nitric acid oceans, and because dissolved ammonium could be incorporated into clays where it would accumulate. If the atmospheric drawdown were sufficiently severe, then it could even result in a global glaciation due to the loss of pressure broadening from nitrogen. Although anthropic reasoning accounts for why such a metabolism never evolved, a more satisfying explanation is that the kinetic barriers to the $N_2$-$O_2$-$H_2O$ reaction are insurmountable. The process by which $N_2$ is converted to nitrate by Earth life is complex with multiple steps: Nitrogen fixers expend energy to convert $N_2$ to $NH_3$ under anaerobic conditions (or with adaptations to overcome high oxygen levels) to overcome an activation energy barrier (*34*), and nitrifiers oxidize $NH_3$ to nitrate under aerobic conditions. Perhaps, the enzymatic machinery required to split the $N_2$ triple bond and combine these two steps under aerobic conditions is too complex and energy-intensive to maintain.

### Practicality of early Earth disequilibrium biosignatures

The most important disequilibrium species in both the Proterozoic ($O_2$, $N_2$, ocean) and the Archean ($CH_4$, $N_2$, $CO_2$, ocean) are, in principle, detectable on exoplanets. In the Archean, high $CH_4$ should be readily detectable (*35*), and $CO_2$ has abundant absorption features (*36*). Nitrogen absorbs at 4.15 μm due to $N_4$, which could be used to infer $N_2$ partial pressure (*37*). Note that it is not necessary to precisely constrain $N_2$ partial pressure to estimate the $CH_4$-$N_2$-$CO_2$-$H_2O$ thermodynamic disequilibrium. For example, for a maximum Archean disequilibrium scenario with very low $N_2$ partial pressure (0.02 bar) and high $CO_2$ (0.95 bar), the available Gibbs energy is 151 J/mol (table S8). So long as there is sufficient $N_2$ (and $CO_2$) to react with $CH_4$ until reaction (4) goes to completion, then a large thermodynamic disequilibrium will exist between $N_2$, CO, $CH_4$, and liquid water. Various techniques have been proposed to detect surface oceans including glint, polarization, and surface mapping [reviewed by Fujii et al. (*38*)] (see section S4 for further discussion).

The remote detectability of oxygen in the Proterozoic atmosphere depends on abundance. If Proterozoic $P_{O_2}$ was <0.1% PAL (present atmospheric level), as has been suggested from one interpretation of Cr isotopes (*14*), then it may not be possible to detect $O_2$ on a "Proterozoic Exo-Earth" with next-generation telescopes (*35*). Extremely low Proterozoic oxygen, however, is difficult to reconcile with photochemical models that show that $O_2$ levels between $10^{-6}$ PAL and 0.1% PAL are unstable against small perturbations in the $O_2$ source flux (*39*), favoring $P_{O_2}$ closer to the upper boundary used for our maximum disequilibrium calculations. In either case, $O_3$, a photochemical product of $O_2$, could be detected even if $O_2$ itself could not (*35*, *40*).

Regardless of uncertainty about precise abundances of $O_2$ and $CH_4$ over Earth history, it is possible to design telescopes capable of constraining the disequilibrium atmospheric constituents described above. A potential problem for quantifying disequilibrium from observations is its sensitivity to variables that are difficult or impossible to observe, such as ocean composition and ocean volume. In the study of Krissansen-Totton et al. (25), we showed the modern Earth's $N_2$-$O_2$-$H_2O$ disequilibrium is relatively insensitive to these variables. The same is true for large Proterozoic disequilibria, which involve the same species.

For the Archean $CO_2$-$N_2$-$CH_4$-$H_2O$ disequilibrium, the equilibrium abundances are insensitive to the unobservable $NH_4^+$ and $HCO_3^-$ ocean molalities. This insensitivity is shown in Fig. 5, which plots the fractional depletion of methane in equilibrium as a function of initial aqueous species molalities. Unless both $NH_4^+$ and $HCO_3^-$ are extremely high, methane should be depleted in equilibrium. Section S4 shows that Archean disequilibrium is robust to uncertainties in ocean composition, volume, temperature, and pressure. The magnitude of the Archean thermodynamic disequilibrium typically varies by less than a factor of 2 across a wide range of assumptions about ocean pH, alkalinity, salinity, volume, temperatures, and pressure.

### Interpretation of early Earth disequilibrium
Would the detection of $CH_4$, $N_2$, $CO_2$, and an $H_2O$ ocean be a reliable exoplanet biosignature? The thermodynamic disequilibria of modern Mars and Titan are overplotted in Fig. 2, and it can be seen that the abiotic photochemical disequilibrium of Mars is comparable in magnitude to the biological disequilibrium of the early Earth. Clearly, the magnitude of atmospheric disequilibrium does not—on its own—indicate the presence of life. Further interpretation is necessary.

In general, atmospheric disequilibria are the product of the generation of free energy balanced by the dissipation of free energy. A large thermodynamic disequilibrium could be the consequence of either a high free energy generation rate or a low dissipation rate (22). For this reason, to evaluate the $CH_4$-$N_2$-$CO_2$-$H_2O$ disequilibrium biosignature, it is necessary to consider the photochemical lifetime of $CH_4$ in such an atmosphere (that is, the rate of free energy dissipation) and possible abiotic sources of $CH_4$ (rate of free energy generation).

Even in reducing atmospheres, $CH_4$ has a geologically short lifetime. Diffusion-limited hydrogen escape would deplete an Earth-like planet of atmospheric methane in ~30 thousand years [(26), p. 215]. Consequently, we would not expect a $CH_4$-$N_2$-$CO_2$-$H_2O$ (liquid) disequilibrium to persist without a substantial flux of $CH_4$ from a planet's surface that is typical of biology. We argue below that large abiotic $CH_4$ fluxes are unlikely and that, where they do occur, they can probably be distinguished by context.

Mantle-derived methane is an implausible abiotic source and could be distinguished by the coexistence of carbon monoxide. Pressure-temperature conditions in the Earth's shallow mantle (<~100 km) strongly favor $CO_2$ over $CH_4$. Any deep mantle $CH_4$ would be converted to $CO_2$ by rapid equilibration long before reaching the surface (41). For terrestrial planets with a more reducing mantle than Earth, significant $CH_4$ outgassing is conceivable. However, a highly reducing mantle would also produce huge CO fluxes (see section S5), and because CO has few abiotic sinks and detectable spectral features (42), mantle-derived methane could readily be identified by coexisting CO. In addition, outgassed CO is unlikely to persist in abundance in the atmospheres of inhabited planets because CO is an excellent source of microbial free energy and carbon (43).

A much discussed scenario for abiotic methane generation is through hydrothermal alteration of crustal mafic rocks (serpentinization), which produces $H_2$, followed by Fischer-Tropsch type (FTT) synthesis. Following the study of Fiebig et al. (44), we estimate the maximum possible abiotic methane flux, $F_{CH_4}$ (mol/year), that could be generated from this process, as follows

$$F_{CH_4} = \frac{P_{Crust}}{M_{FeO}} fr_{FeO} fr_{H_2} fr_{CH_4} \qquad (6)$$

Here, $P_{Crust}$ is the crustal production rate in kilogram per year, $M_{FeO}$ is the molar mass of FeO in kilogram per mole, $fr_{FeO}$ is the weight % (wt %) of FeO in newly produced crust, $fr_{H_2}$ is the maximum fraction of FeO that is converted to $H_2$ by serpentinization reactions, and $fr_{CH_4}$ is the maximum fractional conversion of $H_2$ to $CH_4$ by FTT reactions. Assuming plausible ranges for these unknown variables and sampling their ranges uniformly, we produce a probability distribution for the maximum possible abiotic methane flux.

Today, magma emplacement from ridges, arcs, and plumes is $5.7 \times 10^{13}$ kg/year (45), and with a generous assumption that crustal production may have been 10× higher on the early Earth, we take crustal production, $P_{Crust}$, to range from $5.7 \times 10^{13}$ to $5.7 \times 10^{14}$ kg/year. Whether these high crustal production rates are likely for terrestrial exoplanets is an open question, given that some argue that Archean Earth's crustal production may not have been much greater than modern (46). In addition, it might be possible to put some constraints on exoplanet crustal production rates from observable planetary properties (47).

Hydrothermal alteration is ultimately limited by the availability of FeO. The Earth's basaltic oceanic crust is ~10 wt % FeO, which would imply a modern FeO production rate of $0.1 \times 5.7 \times 10^{13}/(0.056 + 0.016) = 79$ Tmol FeO/year. We will allow fractional FeO content, $fr_{FeO}$, to vary

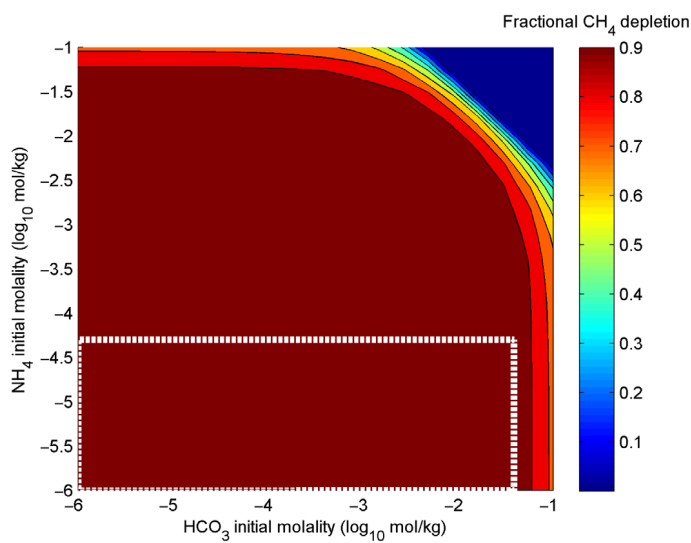

**Fig. 5. Sensitivity of Archean disequilibrium to bicarbonate molality and ammonium molality in ocean, quantities that are probably impossible to directly observe for exoplanets.** Colors shows fraction of methane depleted in equilibrium, as determined by semianalytic calculations. In this case, $P_{CO_2} = 0.49$ bar, $P_{N_2} = 0.5$ bar, and $P_{CH_4} = 0.01$ bar. As can be seen, most parts of parameter space have high $CH_4$ depletion, that is, $CH_4$ is in disequilibrium. Thus, unless both bicarbonate and ammonium molalities are extremely large, detectable quantities of methane are out of equilibrium with an $N_2$-$CO_2$ atmosphere and ocean. The white dashed line box denotes the plausible Archean range.

from 10 to 25 wt %, considering possible Mars- or Moon-like crustal compositions.

A key unknown is what fraction of FeO is oxidized by water to liberate $H_2$, $fr_{H_2}$. A naïve upper estimate would assume that all the crustal FeO is oxidized to yield $H_2$ by the following equation

$$3FeO + H_2O \rightarrow Fe_3O_4 + H_2 \quad (7)$$

Such an assumption is unrealistic, however. On Earth, multiple lines of evidence suggest that the $H_2$ production rate from serpentinization is around 0.2 Tmol/year (48, 49), which is 0.25% of the total FeO production, two orders of magnitude less than a one-third theoretical maximum from the stoichiometry of Eq. 7. We take $fr_{FeO} = 0.0025$ as our lower bound for the fractional conversion of $H_2$ to FeO. On the modern Earth, $H_2$ production is low because only 12% of crustal $Fe^{2+}$ is converted to $Fe^{3+}$ as water cannot permeate all the oceanic crust (45). Furthermore, only 1 to 2% is due to serpentinization; the remaining 10 to 11% is from oxidation by sulfate, which does not generate $H_2$ [(26), chap. 10; (45)]. Most of the $H_2$ produced from serpentinization is derived from ultramafic, slow-spreading crust, which only constitutes 24% of the total crustal production (48). A nonlinear relationship also exists between degree of serpentinization and $H_2$ production, and for <50% serpentinization, negligible $H_2$ is generated (49). For the upper bound, we generously assume that the entirety of an Exo-Earth's FeO oxidation is due to $H_2$-producing serpentinization through Eq. 7, as might occur with more ultramafic crust, giving $fr_{FeO} = 0.12/3 = 0.04$.

The fractional conversion of $H_2$ to $CH_4$, $fr_{CH_4}$, is the final parameter for calculating maximum abiotic $CH_4$ fluxes. For maximum conversion, we assume that all $H_2$ makes $CH_4$ via FTT synthesis

$$CO_2 + 4H_2 \rightarrow 2H_2O + CH_4 \quad (8)$$

That is, we assume $fr_{CH_4} = 0.25$, aware that this is an overestimate. The conversion of $CO_2$ to $CH_4$ by Eq. 8 is thermodynamically favorable in low-temperature hydrothermal systems, but it is unclear whether natural systems can overcome kinetic barriers on a global scale without biological catalysts. Although evidence from field studies suggests that abiotic methane is generated in some hydrothermal systems (44, 50), laboratory experiments typically find very low methane yields from FTT synthesis from olivine. High experimental methane production has been reported (51, 52), but similar experiments with $^{13}C$-labeled carbon have shown that the methane produced is derived from background organic carbon contamination (53, 54). Only very specific laboratory conditions yield high $CH_4$ production from olivine. For example, when pressure is low enough for gas-phase reactions, abiotic $CH_4$ production is high (53), but it is unlikely that gas-phase reactions would occur at great depth in the crust (44). Similarly, the presence of Fe-Ni catalysts enables $CH_4$ production (55), but most of Earth's crust is not sufficiently reducing to have such catalysts (41). Guzmán-Marmolejo et al. (56) argued that $CH_4$ production is further restricted by $CO_2$ availability in the crust, limiting $H_2/CH_4$ ratios to ~13. Higher $CO_2$ concentrations are unlikely to overcome this restriction because $Fe^{2+}$ will be incorporated into siderite rather than form magnetite and $H_2$ (57). Empirically, $H_2/CH_4$ ratios in hydrothermal systems are highly variable (48), so it is clear that the kinetic barriers to $CH_4$ formation and $CO_2$ limitations are generally present.

For our lower bound on $fr_{CH_4}$, we will adopt the empirical average from ultramafic-hosted hydrothermal fluids, $H_2/CH_4 = 12$ [table 1 of Keir (48)], which implies $fr_{CH_4} = 1/16$, although this likely overestimates global methane production for the reasons discussed above.

Figure 6 shows the probability distribution for $F_{CH_4}$ obtained by uniformly sampling our chosen ranges for $P_{Crust}$, $fr_{FeO}$, $fr_{H_2}$, and $fr_{CH_4}$. Both the modern biological methane flux (58) and plausible biological Archean fluxes (59) are much larger than the distribution of maximum abiotic fluxes. On the basis of current understanding, the conditions required to generate large fluxes of abiotic methane are specific and implausible: All unknown variables need to be at the high end of their ranges, and our upper estimate for $fr_{H_2}$ must be an underestimate. However, underestimation of $fr_{H_2}$ is implausible because for fast rates of crustal production, fractional conversion of FeO to $H_2$ will, if anything, be lower than on Earth because water will permeate a smaller fraction of the total crust given faster spreading rates and thicker crust. In addition, the gas-phase reactions required to overcome kinetic barriers to $CH_4$ production will be inhibited at greater crustal thicknesses (44).

Impacts have also been proposed as a source of abiotic methane. During an impact, Fe or Ni catalysts in an asteroid could produce $CH_4$ from CO and $H_2$. Kress and McKay (60) modeled the kinetics of $CH_4$ production during the cooling of an impactor fireball and concluded that this could produce an abiotic methane flux during the late heavy bombardment greater than the modern Earth's biological flux. However, an exceptionally high mass flux is required. Using more plausible mass fluxes, Kasting (61) estimated an abiotic methane flux from impactors to be only ~0.3 Tmol/year at 3.8 Ga. This could potentially be increased to 1 Tmol/year if all impact ejecta are serpentinized. However, this too is unrealistically large because complete 12:1 FeO oxidation to $CH_4$ production is far from guaranteed, as discussed above. In any case, for terrestrial exoplanets, it may be possible to rule out large impactor fluxes from the system age, dust levels, or the absence of transiting planetesimals. For very high impactor fluxes, observable atmospheric consequences such as dust or CO may be detectable.

We conclude that large abiotic sources of $CH_4$ are either improbable or identifiable with other observations. Furthermore, the rich absorption spectra of $CO_2$, $CH_4$, and water vapor mean that they are the most readily detectable gases on an exoplanet with an anoxic atmosphere in the near future. Confirmation of a liquid water ocean and atmospheric $N_2$ that contribute to the full disequilibrium of an Archean Earth twin would require follow-up observations with next-generation space telescopes. However, the detection of $CO_2$ and $CH_4$ alone on a habitable exoplanet is a potential disequilibrium biosignature because carbon is present at the extreme ends of the redox ladder (the +4 and −4 states). This $CO_2$-$CH_4$ disequilibrium pair is more compelling than $CH_4$ alone, which could be primordial from a migrated icy world or outgassed from an extremely reducing mantle. An atmosphere rich in $CO_2$ and $CH_4$ has been proposed to explain a warm early Mars (62), but this atmosphere would be very transient (<1 million years) because $CH_4$ released from clathrates would be rapidly photodissociated.

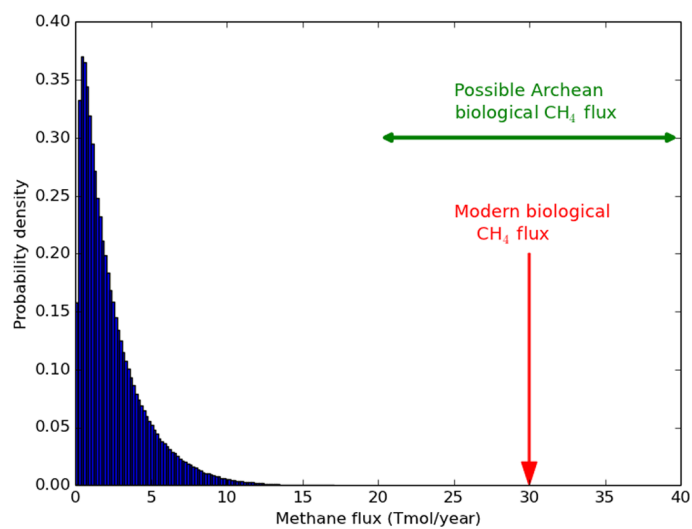

**Fig. 6. Probability distribution for maximum abiotic methane production from serpentinization on Earth-like planets.** This distribution was generated by sampling generous ranges for crustal production rates, FeO wt %, maximum fractional conversion of FeO to $H_2$, and maximum fractional conversion of $H_2$ to $CH_4$, and then calculating the resultant methane flux 1 million times (see the main text). The modern biological flux (58) and plausible biological Archean flux (59) far exceed the maximum possible abiotic flux. These results support the hypothesis that the co-detection of abundant $CH_4$ and $CO_2$ on a habitable exoplanet is a plausible biosignature.

### Kinetic and geological considerations

The mere detection of $CO_2$ and $CH_4$ (and the absence of CO) does not necessarily imply the presence of life. Instead, it is desirable to calculate the necessary methane source flux given observed atmospheric abundances and the stellar spectrum (56). If the inferred source flux is greater than any known plausible abiotic mechanism (Fig. 6), or if plausible abiotic fluxes can be ruled out by contextual information, then life would be left as a reasonable hypothesis. For planets with anoxic atmospheres such as the $CO_2$-$CH_4$-$N_2$ atmospheres considered here, the methane abundance is set by the balance between photochemical destruction of methane with subsequent diffusion-limited hydrogen escape and the methane source flux from the surface (see section S6 for details). Thus, given an observed atmospheric $CH_4$ abundance, it is possible to infer the minimum $CH_4$ flux required to maintain this. Calculations (section S6) suggest that $CH_4$ abundances in excess of $10^{-3}$ imply methane source fluxes in excess of 7 Tmol/year, which is likely biological, whereas $CH_4$ abundances in excess of $10^{-2}$ imply methane source fluxes in excess of 50 Tmol/year, which is very likely biological (compare Fig. 6), when seen in combination with $CO_2$ and the absence of CO. These results are largely independent of stellar type, but more precise photochemical calculations ought to be applied in the future to better estimate implied fluxes from abundances.

The $CO_2$-$CH_4$ pair might be the most easily detectable exoplanet biosignature. For habitable planets around M-dwarfs with Earth-like biogenic fluxes, the transit transmission features from $CO_2$ and $CH_4$ may require shorter integration times to resolve than $O_2$ or $O_3$ features with JWST (63).

The kinetics and geological cycling of atmospheric nitrogen are also worth considering, given their contribution to disequilibria in both anoxic and oxic atmospheres. In (25), we presented a calculation to show that if life disappeared from Earth, lightning would convert atmospheric $N_2$ and $O_2$ to nitrate, depleting atmospheric oxygen in 20 to 200 Ma. In contrast, it has been argued that nitrate may be reduced to ammonia in mid-ocean ridge hydrothermal systems, which can then return to the atmosphere and be photochemically oxidized back to $N_2$ (64). However, the reduction of nitrate will yield ammonium, which is readily sequestered into silicates (65, 66). Nitrogen-bearing crust may then be subducted, where some nitrogen may return to the atmosphere via arc volcanism, but the rest will continue to the mantle because of the stability of ammonium-bearing silicates at high temperatures (67). It is therefore reasonable to expect an $N_2$-$O_2$-$H_2O$ disequilibrium to disappear without the continuing influence of life, although the precise time scale for the depletion of oxygen and nitrogen will depend on geological cycling that is difficult to quantify. Atmospheric $N_2$ persists on Venus because it is in equilibrium with the $CO_2$-dominated atmosphere (25) and there is no mechanism to draw down $N_2$ into the crust because of the lack of water.

### Solid-state disequilibria

We have not included solid states of matter in our equilibrium calculations. In reality, if the modern Earth's atmosphere-ocean system were allowed to relax to equilibrium, then much of the atmospheric $O_2$ would react with the crust via oxidative weathering, and some dissolved carbon may form carbonate-bearing rocks. There is also a large disequilibrium between organic carbon and ferric iron in the crust, both of which have accumulated over time from photosynthesis and hydrogen escape, that far exceeds the disequilibrium in the fluid reservoir (25). Although there are $3.7 \times 10^{19}$ mol $O_2$ in the atmosphere and oceans, there are $5.1 \times 10^{20}$ mol $O_2$ equivalent $Fe^{3+}$ and sulfate in sedimentary rocks, and $\sim 2 \times 10^{21}$ mol $O_2$ equivalent excess $Fe^{3+}$ in igneous and metamorphic rocks (68). These crustal rocks are in disequilibrium with the $<1.3 \times 10^{21}$ mol $O_2$ equivalent reduced carbon in the crust (68). We therefore expect the biogenic disequilibrium in Earth's crustal reservoir to be several orders of magnitude larger than that of the atmosphere-ocean system (25).

However, we chose to ignore solid species because we are interested in remotely detectable disequilibrium biosignatures; detailed crustal compositions cannot be measured for exoplanets, and so, we restrict ourselves to the observable fluid reservoirs. Adding solid phases would not diminish any of the disequilibria described in our analysis but would potentially make the total available energy larger.

One potentially detectable solid-state disequilibrium not considered here is the disequilibrium between atmospheric oxygen and reduced carbon on the surface in the form of biomass [(23), p. 336]. In principle, surface biomass is detectable through the vegetative red edge (16). However, on the modern Earth, even if all the surface biomass ($4.4 \times 10^{17}$ Tmol C) were oxidized, only 1% of atmospheric oxygen would be depleted (69). Biomass oxidation through the following reaction yields 478 kJ/mol

$$CH_2O + O_2 \rightarrow CO_2 + H_2O \qquad (9)$$

Therefore, the available energy from oxidizing all surface biomass is approximately $(4.4 \times 10^{17}$ mol $\times 478$ kJ/mol$)/(1.8 \times 10^{20}$ mol$) = 1200$ J/mol of atmosphere. This dwarfs the $\sim 1$ J/mol from atmospheric methane oxidation but is only half the size of the $N_2$-$O_2$-$H_2O$ disequilibrium on the modern Earth.

The simultaneous detection of atmospheric oxygen and a vegetative red edge would obviously be a more compelling biosignature than oxygen alone. However, we do not account for surface biomass in our disequilibrium through time calculations because quantifying surface organic biomass on exoplanets would be extremely challenging. Tinetti et al. (70) showed that the vegetative red edge is potentially detectable with next-generation direct imaging, though challenging with realistic clouds. However, it is not possible to map a red edge detection to surface biomass because near-infrared reflectance depends on numerous vegetation properties such as leaf thickness and canopy structure (71). Tinetti et al. (70) also considered marine plankton detectability and found that even on planets with shallow oceans and an order of magnitude more plankton than the modern Earth, the disc-integrated spectral features would be weaker than the vegetation signal.

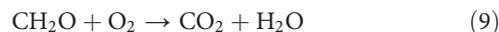

## CONCLUSIONS

Given current knowledge of the evolution of the atmosphere since the Archean, we calculate that Earth's atmosphere has been in thermodynamic chemical disequilibrium since the early Archean as a result of life. The magnitude of this disequilibrium has increased through time, correlated with increases in atmospheric oxygen and probable growth in biomass. The contributions to disequilibrium from solid states of matter were not included in these calculations because our focus is on properties that would be remotely detectable on exoplanets in the future.

In the Proterozoic and Phanerozoic, the coexistence of $N_2$, $O_2$, and liquid $H_2O$ was the largest contributor to chemical disequilibrium. Both $N_2$ and $O_2$ are replenished by biology, and this disequilibrium would not persist in the absence of life because almost all the $O_2$ would be converted to nitric acid in the ocean. A smaller thermodynamic contribution to disequilibrium came from the $CH_4$-$O_2$ couple, which remains a compelling biosignature due to the short kinetic lifetime of $CH_4$ in $O_2$-rich atmospheres.

In the Archean, we show that likely levels of $N_2$, $CH_4$, and $CO_2$ in the presence of liquid $H_2O$ were the largest contributors to disequilibrium. Without life continuously replenishing atmospheric $CH_4$, this disequilibrium would not have persisted because $CH_4$ would have been photolytically destroyed in the upper atmosphere. Large abiogenic fluxes of $CH_4$ (>10 Tmol/year) needed to support high abiogenic $CH_4$ abundances are very unlikely to occur on Earth-like exoplanets, and where they do occur, they can probably be identified through context.

The $CH_4$-$N_2$-$CO_2$-$H_2O$ disequilibrium is thus a potentially detectable biosignature for Earth-like exoplanets with anoxic atmospheres and microbial biospheres. The simultaneous detection of abundant $CH_4$ and $CO_2$ (and the absence of CO) on an ostensibly habitable exoplanet would be strongly suggestive of biology. Specifically, methane mixing ratios $>10^{-3}$ would imply surface fluxes that are potentially biological, whereas mixing ratios $>10^{-2}$ would imply surface fluxes that are likely biological. Biology allows for the coexisting large redox separation of $CH_4$ and $CO_2$ and also readily consumes CO.

## MATERIALS AND METHODS

### Multiphase disequilibrium calculations

The methodology described here updates that used by Krissansen-Totton et al. (25). To calculate thermodynamic equilibrium for multiphase systems, we followed Karpov et al. (72) and used the following expression for the Gibbs energy of a multiphase system (relative to some reference state)

$$\Delta G_{(T,P)} = \sum_i c_i n_i + \sum_\alpha \sum_{i \in \alpha} n_i RT \ln(n_i/n_\alpha) - \sum_{j=\text{aqueous species}} n_j RT \ln(n_w/n_{aq})$$

$$c_i = \begin{cases} \Delta_f G^\circ_{i(T,P)} + RT\ln(\gamma_{fi}) + RT\ln(P), & i \in \text{gas} \\ \Delta_f G^\circ_{i(T,P)} + RT\ln(\gamma_{aw}), & i \in \text{water} \\ \Delta_f G^\circ_{i(T,P)} + RT\ln(\gamma_{ai}) + RT\ln(55.5084), & i \in \text{aqueous} \end{cases}$$

(10)

Here, we have simplified equations in (72) to exclude solid phases and nonwater pure liquids because we do not consider such systems in this study. The variables are defined as follows: $n_i$ is the number of moles of the $i$th species (note that the vector **n** consists of the set of $n_i$); $\Delta_f G^\circ_i(T,P)$ is the standard free energy of formation for the $i$th species at temperature, $T$, and pressure, $P$ [see (25) for explanation of how these were calculated and what databases were used]; α is the index for the phase (gaseous, water, or aqueous); $n_\alpha$ is the total number of moles of species in phase α; $n_w$ is the total number of moles of liquid water in the system; $n_{aq}$ is the total number of moles of aqueous species in the system; $\gamma_{aw}$ is the activity coefficient of water; $\gamma_{ai}$ is the activity coefficient of the $i$th aqueous species; $\gamma_{fi}$ is the fugacity coefficient of the $i$th gaseous species; $R$ is the universal gas constant; $T$ is the temperature of the system (K), a constant; and $P$ is the pressure of the system (bar), a constant.

To calculate the equilibrium state of the Earth's atmosphere-ocean system, we minimize Eq. 10 subject to the constraint that atoms and charge are conserved, where the latter means that aqueous systems are electroneutral (25).

Several improvements to our multiphase equilibrium calculations have been made from the version described by Krissansen-Totton et al. (25). Most importantly, activity coefficients for all aqueous species are calculated using the Pitzer equations rather than the Truesdell-Jones equation. This proved to be important for accurately capturing the Gibbs energy changes for Archean-like atmosphere-ocean systems. Activity coefficients for cations, $M$, and anions, $X$, were specified by the following equations (73)

$$\ln(\gamma_M) = z_M^2 F + \sum_{\text{all anions}} m_a(2B_{Ma} + ZC_{Ma})$$
$$\ln(\gamma_X) = z_X^2 F + \sum_{\text{all cations}} m_c(2B_{cX} + ZC_{cX})$$

where

$$F = -A_\phi \left[ \frac{I^{0.5}}{1+bI^{0.5}} + \frac{2}{b}\ln(1+bI^{0.5}) \right] + \sum_{\text{all pairs}} m_a m_c B'_{Ma}$$
$$Z = \sum_i m_i |z_i|$$
$$B_{MX} = B_{MX}^{(0)} + B_{MX}^{(1)} f(\alpha_1 I^{1/2}) + B_{MX}^{(2)} f(\alpha_2 I^{1/2})$$
$$f(x) = \frac{2[1-(1+x)\exp(-x)]}{x^2}$$
$$B'_{MX} = \frac{B_{MX}^{(1)} f'(\alpha_1 I^{1/2})}{I} + \frac{B_{MX}^{(2)} f'(\alpha_2 I^{1/2})}{I}$$
$$f'(x) = \frac{-2[1-(1+x+x^2/2)\exp(-x)]}{x^2}$$

(11)

Here, the variables are defined as follows: $m_i$ is the molality of the $i$th aqueous species; $m_a$ is the molality of the anion; $m_c$ is the molality of the cation; $z_i$ is the charge of the $z$th aqueous species; $I$ is the ionic strength of the solution, $I = 0.5\sum_i m_i z_i^2$; $\alpha_1 = 2.0$ kg$^{0.5}$mol$^{-0.5}$, $\alpha_2 = 0$ kg$^{0.5}$mol$^{-0.5}$ for all binary systems except 2:2 electrolytes; and $\alpha_1 = 1.4$ kg$^{0.5}$mol$^{-0.5}$, $\alpha_2 = 12$ kg$^{0.5}$mol$^{-0.5}$ for 2:2 electrolytes; $b = 1.2$ kg$^{0.5}$mol$^{-0.5}$ and $A_\phi = 0.3915$ kg$^{0.5}$mol$^{-0.5}$ are constants.

$B_{MX}^{(0)}$, $B_{MX}^{(1)}$, $B_{MX}^{(2)}$, $C_{MX}$ are species-specific binary interaction parameters that were obtained from Appelo and Postma (74) and Marion (75). Note that we have adopted a simplified version of the Pitzer equations by ignoring cation-cation and anion-anion interactions, neutral solute parameters, and triple particle parameters. The excellent agreement between MATLAB and Aspen calculations confirms that neglecting these terms is a reasonable approximation. Methods for calculating fugacity coefficients and the activity of water are the same as those by Krissansen-Totton et al. (25).

## SUPPLEMENTARY MATERIALS

Supplementary material for this article is available at http://advances.sciencemag.org/cgi/content/full/4/1/eaao5747/DC1
Supplementary Text
section S1. The omission of solids, redox-sensitive nonvolatile aqueous species, and ocean heterogeneity
section S2. Full results and Aspen Plus validation
section S3. Reactions associated with Precambrian disequilibria and their Gibbs energy contributions
section S4. Sensitivity of available Gibbs energy to difficult-to-observe variables
section S5. Abiotic $CH_4$ formation from high-temperature processes
section S6. Atmospheric kinetics of methane destruction
fig. S1. Atmosphere-ocean disequilibrium in the Proterozoic (minimum disequilibrium scenario).
fig. S2. Atmosphere-ocean disequilibrium in the Archean (minimum disequilibrium scenario).
fig. S3. Relationship between methane fluxes and atmospheric abundances.
table S1. Proterozoic maximum disequilibrium.
table S2. Proterozoic minimum disequilibrium.
table S3. Proterozoic disequilibrium with 2% PAL of $O_2$.
table S4. Archean maximum disequilibrium.
table S5. Archean minimum disequilibrium.
table S6. Reactions contributing to Proterozoic disequilibrium.
table S7. Reactions contributing to Archean disequilibrium.
table S8. Sensitivity of Archean disequilibrium to difficult-to-observe variables.
References (92–105)

**Acknowledgments:** We thank J. Lustig-Yaeger for helpful discussions. We also thank A. Kleidon, N. Sleep, and the anonymous reviewer for numerous comments that greatly improved the manuscript. **Funding:** This work was supported by the NASA Astrobiology Institute's Virtual Planetary Laboratory (grant NNA13AA93A) and the NASA Exobiology Program (grant NNX15AL23G) awarded to D.C.C. J.K.-T. is supported by NASA Headquarters under the NASA Earth and Space Science Fellowship program (grant NNX15AR63H). **Author contributions:** D.C.C. conceived the project. J.K.-T. and D.C.C. contributed to creating the code and performing the analysis. S.O. compiled estimates of the early Earth's atmosphere and ocean composition. All authors contributed to the drafting of the manuscript. **Competing interests:** The authors declare that they have no competing interests. **Data and materials availability:** All data needed to evaluate the conclusions in the paper are present in the paper and/or the Supplementary Materials. Our MATLAB code for calculating multiphase equilibrium is available on the website of the lead author.

Submitted 17 August 2017
Accepted 19 December 2017
Published 24 January 2018
10.1126/sciadv.aao5747

**Citation:** J. Krissansen-Totton, S. Olson, D. C. Catling, Disequilibrium biosignatures over Earth history and implications for detecting exoplanet life. *Sci. Adv.* **4**, eaao5747 (2018).

**Disequilibrium biosignatures over Earth history and implications for detecting exoplanet life**

Joshua Krissansen-Totton, Stephanie Olson and David C. Catling

**This PDF file includes:**

- Supplementary Text
- section S1. The omission of solids, redox-sensitive nonvolatile aqueous species, and ocean heterogeneity
- section S2. Full results and Aspen Plus validation
- section S3. Reactions associated with Precambrian disequilibria and their Gibbs energy contributions
- section S4. Sensitivity of available Gibbs energy to difficult-to-observe variables
- section S5. Abiotic $CH_4$ formation from high-temperature processes
- section S6. Atmospheric kinetics of methane destruction
- fig. S1. Atmosphere-ocean disequilibrium in the Proterozoic (minimum disequilibrium scenario).
- fig. S2. Atmosphere-ocean disequilibrium in the Archean (minimum disequilibrium scenario).
- fig. S3. Relationship between methane fluxes and atmospheric abundances.
- table S1. Proterozoic maximum disequilibrium.
- table S2. Proterozoic minimum disequilibrium.
- table S3. Proterozoic disequilibrium with 2% PAL of $O_2$.
- table S4. Archean maximum disequilibrium.
- table S5. Archean minimum disequilibrium.
- table S6. Reactions contributing to Proterozoic disequilibrium.
- table S7. Reactions contributing to Archean disequilibrium.
- table S8. Sensitivity of Archean disequilibrium to difficult-to-observe variables.
- References (*92–105*)

# Supplementary Materials

**Supplementary Text**

**section S1. The omission of solids, redox-sensitive nonvolatile aqueous species, and ocean heterogeneity**

We neglected redox-sensitive, non-volatile aqueous species because they are not remotely observable, but this omission does not affect our results significantly. The most abundant redox-sensitive, non-volatile aqueous species in the Precambrian ocean was iron. If the Precambrian ocean was mostly ferruginous, then aqueous $Fe^{2+}$ could react with $O_2$ or with aqueous $NO_3^-$ derived from $N_2$ oxidation (92), thereby adding to the maximum Proterozoic available energy. To test this, we repeated maximum Proterozoic disequilibrium calculations using Aspen Plus and varied the concentration of ferrous iron (with zero initial ferric iron). If $Fe^{2+}$ concentrations in the Precambrian oceans were $7\times10^{-5}$ mol/kg ($5.5\times10^{-4}$ mol per mol of atmosphere), assuming siderite saturation (93), then only ~13 J/mol are added from $Fe^{2+}$ oxidation, a 2% increase from the nominal maximum Proterozoic case. Even if $Fe^{2+}$ concentrations were extremely high due to siderite supersaturation (94), for instance $3\times10^{-3}$ mol/kg (0.023 mol per mol of atmosphere), then the available energy is 1197 J/mol. This is still less than the Phanerozoic disequilibrium, and does not change the qualitative disequilibrium evolution in Fig. 2.

By assuming saturation to obtain initial dissolved gas abundances, we are neglecting the possible contribution from ocean heterogeneity. Specifically, by adopting Henry's law to calculate the initial dissolved oxygen abundance, we may be overestimating the Proterozoic disequilibrium because the Proterozoic ocean was mostly anoxic. However, even at saturation, the number of moles of oxygen in the ocean is about two orders of magnitude less than the atmosphere, and so the total available energy will be overestimated only very slightly. If the maximum Proterozoic disequilibrium calculation is repeated with zero dissolved oxygen in the initial state, then the total available energy is only 6 J/mol lower than when oxygen saturation is assumed (<1% difference). This calculation also shows that oxygen oases in the Archean (95) and Proterozoic (96) will have a negligible effect on the global disequilibrium – a realistic Precambrian ocean with 1-10% oxic waters will yield an available energy somewhere between our $O_2$-saturated nominal calculation, and the zero dissolved oxygen endmembers. However, oxygen oases still represent an important example of local disequilibria maintained by life that would be very difficult to detect remotely.

**section S2. Full results and Aspen Plus validation**

Tables S1, S2, S3, S4, and S5 show full multiphase equilibrium results for our Proterozoic maximum, Proterozoic minimum, Proterozoic mid-level oxygen, Archean maximum, and Archean minimum scenarios, respectively.

**section S3. Reactions associated with Precambrian disequilibria and their Gibbs energy contributions**

Table S6 shows the various reactions contributing to Proterozoic disequilibrium, and the Gibbs energy contribution from each of these reactions. Since methane levels were potentially high in the Proterozoic, the contribution from methane oxidation (75 J/mol) to the maximum Proterozoic disequilibrium is much higher than the ~1.3 J/mol contribution to the modern Earth. Contributions from the oxidation of aqueous species such as ammonium and sulfide only contribute 14% to the total available Gibbs energy (maximum case), and are not relevant for life detection on exoplanets since changes in ocean speciation are not remotely observable.

Table S7 shows the various reactions contributing to Archean disequilibrium, and the Gibbs energy contribution from each of these reactions. The second largest contributor to maximum Archean disequilibrium (after ammonium formation from methane oxidation) is the oxidation of carbon monoxide.

**section S4. Sensitivity of available Gibbs energy to difficult-to-observe variables**

Table S8 shows the thermodynamic disequilibrium for our maximum Archean case where we have varied parameters such as temperature, ocean pH/alkalinity, ocean salinity, ocean volume, atmospheric pressure, and atmospheric bulk composition. Theses variables are challenging to observe for exoplanets, and so this table explores whether it would be possible to quantify the presence of thermodynamic disequilibrium for an Archean-like exoplanet. Results are reported for both Matlab and ASPEN calculations, and the two computational methods agree to within a few percent in every case, validating our Matlab calculations.

The $CH_4$-depleting reaction went to completion (i.e. all $CH_4$ was used up in a forward reaction) regardless of temperature.

Ocean alkalinity was varied by solving the carbonate system equilibria for a system with $pCO_2$ = 0.5 bar and the specified alkalinity from the table S8 (25). Next, Na(+) abundances were adjusted from their nominal values to maintain charge balance. We find that for the high alkalinity case (200 mmol/kg) the $CH_4$-depleting reaction only depletes 89% of the methane in equilibrium, as opposed to over 99% in other sensitivity tests. This case is highlighted in bold in table S8. However, alkalinities this high are unlikely to occur given buffering from silicate weathering (29).

Ocean salinity was varied by increasing (or decreasing) Na(+) and Cl(-) by the same amount, thereby preserving carbonate alkalinity. The $CH_4$-depleting reaction went to completion regardless of salinity.

Ocean volume was varied by scaling the liquid water abundances and other aqueous constituents by the specified amount. All reactions went to completion except the very low ocean volume case, which is highlighted in bold in table S8. Note that in the extreme case of no ocean, reaction (4) does not occur and the Archean disequilibrium is small (Fig. 2) and is dominated by CO oxidation (table S7). Promising strategies for constraining ocean volume remotely such as glint, polarization, time-resolved photometry, geophysical theory, and thermal inertia, are discussed in Krissansen-Totton *et al.* (25). However, even if surface oceans cannot be directly detected, for instance because of haze obscuring surface features, then it may still be possible to infer the presence of an ocean. If water vapor abundance at a particular pressure level can be constrained, then climate models could be applied to check for consistency with a steam atmosphere.

Archean total pressure may have been less than 1 bar (97-99), and exoplanet surface pressures may vary considerably, and so we also calculate the sensitivity of our disequilibrium to total atmospheric pressure. Atmospheric pressure was varied by changing the gas-phase pressure and adjusting the liquid-to-gas ratio by the inverse amount. This accounts for Archean pressure being lower (higher) because of fewer (more) moles in the atmosphere, not because of a change in gravity. The carbonate equilibria was solved for the new initial conditions, and Na(+) abundances were adjusted to preserve charge balance. For the more extreme cases, the initial water phase separation needed to be readjusted using Henry's law (see methods). The $CH_4$-depleting reaction went to completion regardless of pressure, and for modest changes in pressure (0.5 bar or 2 bar), the available Gibbs energy is very similar to the 1 bar case.

Finally, we considered lower pressure in conjunction with different bulk abundances. This is relevant to the early Archean where there is considerable uncertainty in atmospheric composition. The maximum Archean disequilibrium was calculated for both 17 % $N_2$ with 80% $CO_2$, and for 77% $N_2$ with 20% $CO_2$ (the remaining 3% is water vapor and $CH_4$ as shown in table S4). The available energy in these cases is very similar to the nominal case (47% $CO_2$ and 50% $N_2$). Even with only 2% $N_2$ abundance and 95% $CO_2$ abundance (at 1 bar), the available energy is still comparable to the nominal case. These results imply that it is not necessary to constrain the $N_2$ abundance precisely in order to calculate thermodynamic disequilibrium. So long as there is enough $N_2$ to oxidize all the $CH_4$, reaction (4) will go to completion and the available energy will be large.

**section S5. Abiotic $CH_4$ formation from high-temperature processes**

Abiotic $CH_4$ outgassed from a highly reduced mantle would be easily distinguishable from biotic methane. The thermodynamic calculations reported here are derived in Catling and Kasting (26, Ch. 7). For temperature-pressure conditions relevant to surface volcanism (T = 1200 °C, P = 5 atm) and Earth-like oxygen fugacity for the upper mantle (near the quartz-fayalite-magnetite buffer, $f_{O_2} = 10^{-8.5}$ bar), the predicted $CH_4/CO_2$ ratio of magmatic gases in thermodynamic equilibrium is $9.4 \times 10^{-11}$. This result explains why negligible $CH_4$ is outgassed from volcanoes on the modern Earth (100), except from the thermal decomposition of organic matter.

Let us consider scenarios that might make volcanic $CH_4$ non-negligible. If the mantle oxygen fugacity is four orders of magnitude lower than Earth's upper-mantle, comparable to inferences from some Mars meteorites (101), then the predicted $CH_4/CO_2$ ratio is ~0.01. The $CO_2$ outgassing flux on the modern Earth is around 10 Tmol/yr (102), and so even in this extreme case the predicted methane flux is only 0.1 Tmol/yr. However, if the total outgassing flux was 1-2 orders of magnitude higher, as is possible for the early Earth, then the volcanic methane flux could extend into the Tmol/yr range, comparable to the modern biological flux. Yet in this scenario, the predicted $CO/CH_4$ ratio is ~6. The atmosphere of such a planet would be rich in

CO. Because CO has many readily observable spectral features (42), this planet would be easily distinguished from a $CO_2$-$CH_4$ biosignature. The coexistence of atmospheric carbon at opposite ends of the redox spectrum (+4 in $CO_2$ and -4 in $CH_4$) with no intermediate (+2) CO cannot be explained by high-temperature outgassing alone.

Other high temperature processes have been proposed for abiotic $CH_4$ such as iron carbonate decomposition, graphite metamorphism, clay-catalyzed hydrocarbon synthesis, and organosulfur reactions (41,103). These pathways are either unlikely to be significant on a global scale, have not been well quantified, or in the case of graphite, are likely ultimately sourced from organic carbon.

**section S6. Atmospheric kinetics of methane destruction**

Figure S3 shows how surface methane fluxes are related to steady state methane abundances for anoxic atmospheres. The solid black line is from Kasting and Brown (77) where a Sun-like star is assumed and a photochemical model is used to compute steady state $CH_4$ abundances in Earth's early atmosphere. The relationship looks similar for an Archean-Earth-like exoplanet atmosphere around M-dwarfs because the dominant sink for atmospheric methane in anoxic atmospheres is photochemical destruction and hydrogen escape (104). In steady state, the net H source to the anoxic atmosphere will be balanced by the net loss of H via escape at the top of the atmosphere

$$F_{escape} = F_{input} \text{ (of H atoms)} \tag{S1}$$

If we assume the H input source is primarily the $CH_4$ flux into the atmosphere, as is likely for planets with large biological $CH_4$ fluxes (59), and that H escape is diffusion limited, then equation (S1) becomes

$$k_{esc} f_{tot} = 4 F_{CH_4} \tag{S2}$$

where $k_{esc}$=2.5×10$^{13}$ cm$^{-2}$ s$^{-1}$ is from diffusion-limited escape theory (see below), $F_{CH_4}$ is the $CH_4$ flux to the atmosphere (Tmol $CH_4$/yr), and we weight $CH_4$ fluxes by the number of H atoms per molecule.

Because of efficient turbulent mixing below the homopause, the total hydrogen mixing ratio at the homopause is the same as that immediately above the troposphere and dominated by $CH_4$, so

$$f_{tot} \approx 4 f_{CH_4} \tag{S3}$$

Where $f_{CH_4}$ is the $CH_4$ mixing ratio. Thus by combining equations (S2) and (S3) and rearranging with a conversion factor to global Teramole (10$^{12}$ mole) fluxes of $CH_4$, we have

$$f_{CH_4} = F_{CH_4} / k_{esc} = F_{CH_4} \text{ (Tmol/yr)} / 6680 \text{ (Tmol/yr)} \qquad (S4)$$

The constant $k_{esc} = b(T)/H_a(T)$, where $b(T)$ is a binary diffusion coefficient weakly dependent on the temperature at the homopause and $H_a(T)$ is the temperature-dependent scale height at the homopause (see (26) for details). Here, we assume $b(T=208\text{ K})/H_a(T=208\text{ K}) = 2.5 \times 10^{13}$ molecules cm$^{-1}$ s$^{-1}$ = 6680 Tmol/yr.

In fig. S3 we plot the steady state abundances derived from this diffusion-limited escape rate for homopause temperatures of T=150 K, 200 K, and 250 K, and an Earth-like scale-height (corrected for temperature). These lines provide a lower limit on CH$_4$ fluxes for any given CH$_4$ abundance. For instance, if there are other H-bearing gases (e.g. high H$_2$), then $f_{tot} > 4 f_{CH_4}$ and the actual CH$_4$ flux required to maintain an observed CH$_4$ abundance may be higher than the diffusion limit suggests (note that a change in this direction would only strengthen the case for biology). For planets with different atmospheric bulk abundances (e.g. CO$_2$-dominated) or different surface gravities, the homopause scale height will change and so the diffusion limit will be slightly shifted relative to the lines plotted in fig. S3.

In principle, the diffusion limit is independent of stellar type, and so the linear relationship between fluxes and abundances in fig. S17 is expected to apply to all habitable exoplanets with anoxic atmospheres. Escape rates slower than the diffusion limit are unlikely because even for cold thermospheres, non-thermal escape process compensate for lowered Jeans' escape (68). Photodissociation of CH$_4$ and therefore the rate of hydrogen escape is ultimately limited by the shortwave (<145 nm) stellar flux. The dominant stellar flux at these wavelengths comes from the Lyman-α line (121.6 nm), and habitable zone planets around M stars typically receive higher Lyman-α fluxes than habitable zones planets around G stars (105). We therefore expect hydrogen escape around M stars to be limited by diffusion, and not by CH$_4$ photodissociation.

Figure S3 suggests that CH$_4$ abundances in excess of $1.5 \times 10^{-3}$ imply surface fluxes in excess of 10 Tmol CH$_4$/yr, which is likely biological (Fig. 6), whereas CH$_4$ abundances in excess of $10^{-2}$ imply surface fluxes in excess of 50 Tmol CH$_4$/yr, which is very likely biological (Fig. 6). Today's flux of CH$_4$ to the Earth's atmosphere is ~30 Tmol CH$_4$/yr, for comparison.

Naturally, the calculations outlined above are to be interpreted as approximations. If methane is observed on a nearby exoplanet, then photochemical modeling using the observed stellar spectrum might be attempted to produce a more precise version of fig. S3. Once CH$_4$ abundances have been constrained by observation, the corresponding escape flux, and by extension the surface source flux can be inferred. This can then be compared to the likely distribution of maximum abiotic source fluxes.

*Supplementary figures and tables*

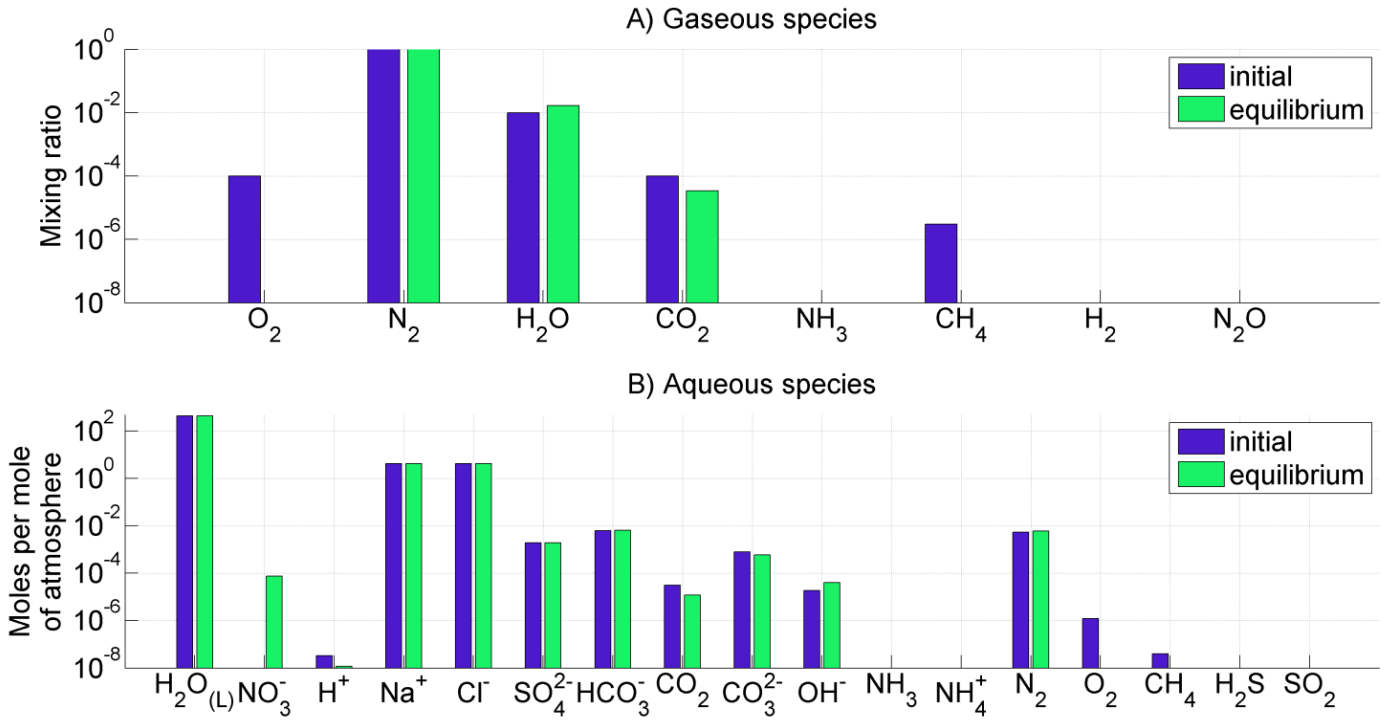

**fig. S1. Atmosphere-ocean disequilibrium in the Proterozoic (minimum disequilibrium scenario).** Blue bars denote assumed initial abundances from the literature, and green bars denote equilibrium abundances calculated using Gibbs free energy minimization. Subplots separate (**A**) atmospheric species and (**B**) ocean species.

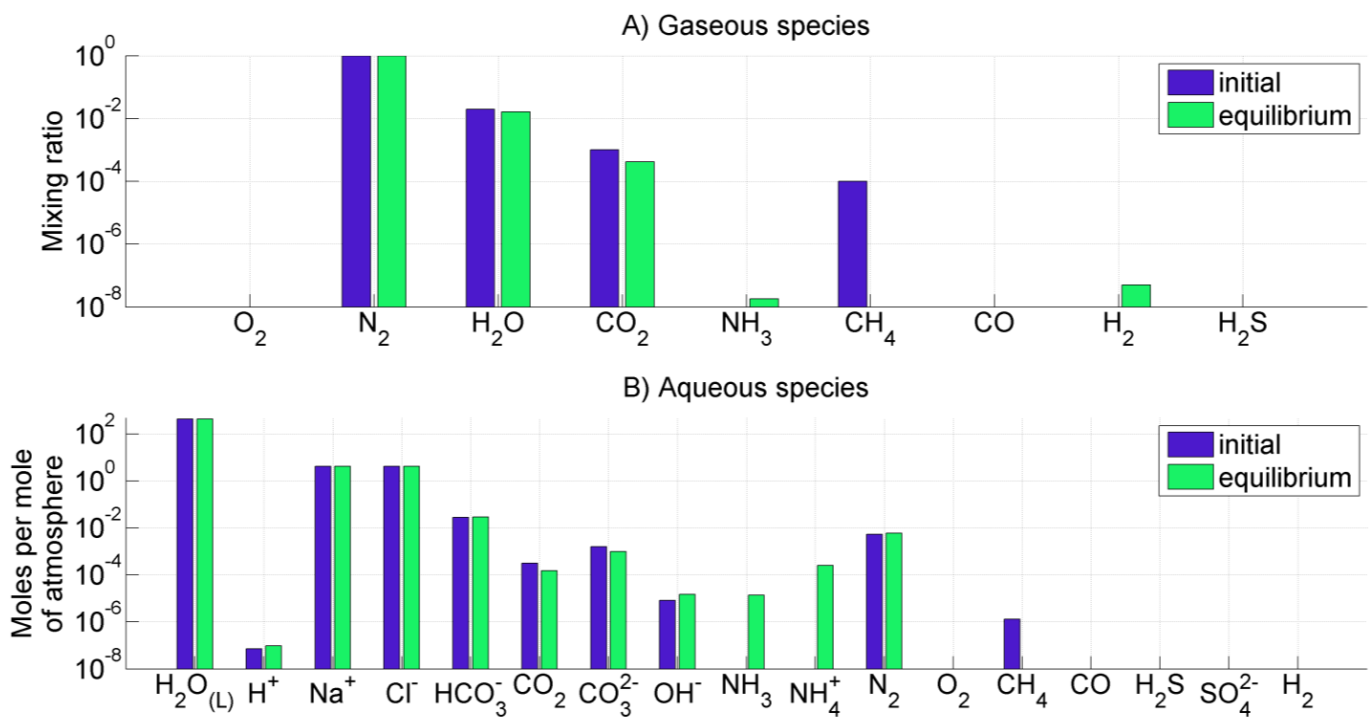

**fig. S2. Atmosphere-ocean disequilibrium in the Archean (minimum disequilibrium scenario).** Blue bars denote assumed initial abundances from the literature, and green bars denote equilibrium abundances calculated using Gibbs free energy minimization. Subplots separate (**A**) atmospheric species and (**B**) ocean species. Note that even in the minimum disequilibrium case, atmospheric $CH_4$ is depleted by reaction to equilibrium.

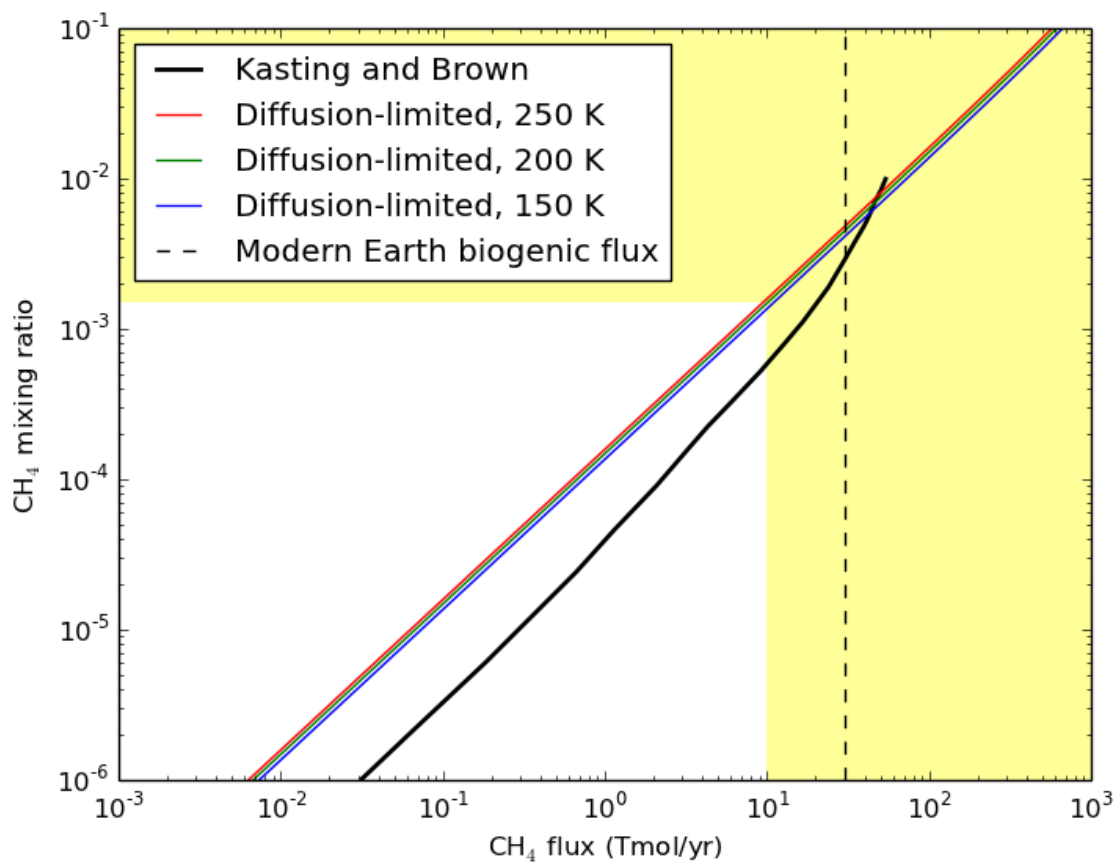

**fig. S3. Relationship between methane fluxes and atmospheric abundances.** The black line is derived in (77) using a photochemical model for the early Earth's atmosphere, assuming a Sun-like star. The red, green, and blue lines are derived assuming diffusion-limited escape of hydrogen with homopause temperatures of 250 K, 200 K, and 150 K, respectively. Diffusion-limited escape demarcates the minimum methane flux required to sustain any given methane abundance. Note that the diffusion-limited escape model may overestimate the $CH_4$ mixing ratio at any given flux by a factor of 2-3 because it does not account for other $CH_4$ sinks.

**table S1. Proterozoic maximum disequilibrium.** Mixing ratios are reported for gaseous species, and moles per mole of atmosphere for aqueous species. Note that final Aspen values that do not completely conserve mass and charge are an artifact of Aspen automatically adjusting initial abundances slightly before performing Gibbs energy calculations.

|  | Initial | Matlab final | Aspen final |
|---|---|---|---|
| $H_2O$(liquid) | 436.788155 | 436.793984 | 436.7938 |
| $O_2$ | 0.03 | 2.64E-05 | 5.59E-05 |
| $N_2$ | 0.86 | 0.84759313 | 0.8485178 |
| $H_2O$ | 0.01 | 0.01627808 | 0.0163399 |
| $CO_2$ | 0.1 | 0.11308654 | 0.1148073 |
| $NH_3$ | 1.00E-15 | 3.47E-12 | 0 |
| $CH_4$ | 0.0001 | 1.29E-12 | 0 |
| $H_2$ | 2.00E-06 | 5.32E-12 | 0 |
| $N_2O$ | 1.00E-06 | 1.47E-11 | 0 |
| $NO_3(-)$ | 7.83E-15 | 0.02387899 | 0.0238104 |
| $H(+)$ | 8.74E-06 | 0.001014 | 0.00119341 |
| $Na(+)$ | 4.3773073 | 4.3773073 | 4.338146 |
| $Cl(-)$ | 4.27587006 | 4.27587006 | 4.237616 |
| $SO_4(2-)$ | 0.03916493 | 0.03916493 | 0.0388458 |
| $HCO_3(-)$ | 0.02348567 | 0.00024239 | 0.000221296 |
| $CO_2(aq)$ | 0.03135339 | 0.04162247 | 0.0397126 |
| $CO_3(2-)$ | 1.10E-05 | 9.20E-10 | 3.38E-10 |
| $OH(-)$ | 7.02E-08 | 1.00E-15 | 4.13E-10 |
| $NH_3(aq)$ | 7.79E-13 | 3.89E-12 | 0 |
| $NH_4(+)$ | 0.00039165 | 4.61E-12 | 0 |
| $N_2(aq)$ | 0.00470212 | 0.00536632 | 0.0044815 |
| $O_2(aq)$ | 0.00037233 | 3.37E-07 | 5.91E-07 |
| $CH_4(aq)$ | 1.32E-06 | 1.28E-12 | 0 |
| $H_2S(aq)$ | 1.00E-15 | 1.52E-12 | 0 |
| $SO_2(aq)$ | 7.83E-15 | 4.58E-12 | 0 |
| **Available Gibbs energy, $\Phi$ (J/mol)** |  | **884** | **860** |

**table S2. Proterozoic minimum disequilibrium.** Mixing ratios are reported for gaseous species, and moles per mole of atmosphere for aqueous species. Note that final Aspen values that do not completely conserve mass and charge are an artifact of Aspen automatically adjusting initial abundances slightly before performing Gibbs energy calculations.

|  | Initial | Matlab final | Aspen final |
|---|---|---|---|
| $H_2O$(liquid) | 436.7881549 | 436.7814412 | 436.7812 |
| $O_2$ | 0.0001 | 2.64E-12 | 0 |
| $N_2$ | 0.99 | 0.989253694 | 0.9894527 |
| $H_2O$ | 0.01 | 0.01657188 | 0.0167461 |
| $CO_2$ | 0.0001 | 3.39E-05 | 1.25128E-05 |
| $NH_3$ | 1.00E-15 | 4.09E-12 | 0 |
| $CH_4$ | 3.00E-06 | 1.44E-12 | 0 |
| $H_2$ | 1.00E-15 | 1.45E-11 | 0 |
| $N_2O$ | 1.00E-15 | 1.33E-11 | 0 |
| $NO_3(-)$ | 7.83E-15 | 7.61E-05 | 7.61295E-05 |
| $H(+)$ | 3.28E-08 | 1.20E-08 | 1.56172E-08 |
| $Na(+)$ | 4.287638267 | 4.287638267 | 4.28568 |
| $Cl(-)$ | 4.275870063 | 4.275870063 | 4.273917 |
| $SO_4(2-)$ | 0.001958246 | 0.001958246 | 0.00195735 |
| $HCO_3(-)$ | 0.006265174 | 0.00653947 | 0.00658799 |
| $CO_2$(aq) | 3.14E-05 | 1.21E-05 | 4.57563E-06 |
| $CO_3(2-)$ | 7.84E-04 | 5.98E-04 | 5.75E-04 |
| $OH(-)$ | 1.87E-05 | 4.02E-05 | 3.36E-05 |
| $NH_3$(aq) | 7.79E-13 | 4.40E-12 | 0 |
| $NH_4(+)$ | 7.83E-15 | 4.62E-12 | 0 |
| $N_2$(aq) | 0.005412906 | 0.006121147 | 0.00592217 |
| $O_2$(aq) | 1.24E-06 | 7.42E-11 | 0 |
| $CH_4$(aq) | 3.96E-08 | 1.42E-12 | 0 |
| $H_2S$(aq) | 7.83E-15 | 1.46E-12 | 0 |
| $SO_2$(aq) | 7.83E-15 | 4.08E-12 | 0 |
| **Available Gibbs energy, $\Phi$ (J/mol)** |  | **9.54** | **9.26** |

**table S3. Proterozoic disequilibrium with 2% PAL of O₂.** Mixing ratios are reported for gaseous species, and moles per mole of atmosphere for aqueous species. The available energy from this scenario separates the lightly and darkly shaded region in Fig. 2. Note that final Aspen values that do not completely conserve mass and charge are an artifact of Aspen automatically adjusting initial abundances slightly before performing Gibbs energy calculations.

|  | Initial | final matlab | final - ASPEN |
|---|---|---|---|
| $H_2O$(liquid) | 436.7882 | 436.7825 | 436.7822 |
| $O_2$ | 0.004 | 1.00E-15 | 3.55854E-08 |
| $N_2$ | 0.985897 | 0.983636 | 0.9887316 |
| $H_2O$ | 0.01 | 0.016433 | 0.0167926 |
| $CO_2$ | 0.0001 | 0.001867 | 0.0019125 |
| $NH_3$ | 1.00E-15 | 7.77E-13 | 0 |
| $CH_4$ | 3.00E-06 | 5.83E-14 | 0 |
| $H_2$ | 1.00E-15 | 1.27E-12 | 0 |
| $N_2O$ | 1.00E-15 | 3.27E-12 | 0 |
| $NO_3(-)$ | 7.83E-15 | 0.003196 | 0.0031961 |
| $H(+)$ | 3.28E-08 | 8.36E-07 | 9.05431E-07 |
| $Na(+)$ | 4.287638 | 4.287638 | 4.28568 |
| $Cl(-)$ | 4.27587 | 4.27587 | 4.273917 |
| $SO_4(2-)$ | 0.001958 | 0.001958 | 0.00195735 |
| $HCO_3(-)$ | 0.006265 | 0.004645 | 0.00463422 |
| $CO_2$(aq) | 3.14E-05 | 0.000666 | 0.000624464 |
| $CO_3(2-)$ | 7.84E-04 | 5.57E-06 | 9.08882E-06 |
| $OH(-)$ | 1.87E-05 | 5.25E-07 | 5.30271E-07 |
| $NH_3$(aq) | 7.79E-13 | 8.48E-13 | 0 |
| $NH_4(+)$ | 7.83E-15 | 9.07E-13 | 0 |
| $N_2$(aq) | 0.005413 | 0.006076 | 0.00508327 |
| $O_2$(aq) | 1.24E-06 | 7.14E-10 | 3.6597E-10 |
| $CH_4$(aq) | 3.96E-08 | 2.77E-13 | 0 |
| $H_2S$(aq) | 7.83E-15 | 2.79E-13 | 0 |
| $SO_2$(aq) | 7.83E-15 | 8.60E-13 | 0 |
| **Available Gibbs energy, $\Phi$ (J/mol)** |  | **135** | **128** |

**table S4. Archean maximum disequilibrium.** Mixing ratios are reported for gaseous species, and moles per mole of atmosphere for aqueous species. Note that final Aspen values that do not completely conserve mass and charge are an artifact of Aspen automatically adjusting initial abundances slightly before performing Gibbs energy calculations.

|  | Initial | final matlab | final - ASPEN |
|---|---|---|---|
| $H_2O$(liquid) | 436.77573 | 436.734522 | 436.9592 |
| $O_2$ | 2.00E-07 | 1.63E-13 | 0 |
| $N_2$ | 0.5 | 0.48682701 | 0.4874388 |
| $H_2O$ | 2.00E-02 | 0.01591615 | 0.0158566 |
| $CO_2$ | 0.47 | 0.44878826 | 0.4465212 |
| $NH_3$ | 1.00E-09 | 1.59E-08 | 1.97E-08 |
| $CH_4$ | 0.01 | 1.91E-05 | 3.48E-05 |
| CO | 0.001 | 1.00E-15 | 1.08E-11 |
| $H_2$ | 0.0001 | 5.98E-08 | 7.05E-08 |
| $H_2S$ | 3.1326E-05 | 0.0004157 | 0.00047851 |
| $H(+)$ | 3.29E-06 | 3.02E-06 | 3.16E-06 |
| $Na(+)$ | 4.59192496 | 4.59192496 | 4.594223 |
| $Cl(-)$ | 4.28E+00 | 4.27587006 | 4.277862 |
| $HCO_3(-)$ | 0.31253612 | 0.3402181 | 0.3402828 |
| $CO_2$(aq) | 0.16551098 | 0.17043556 | 0.172843 |
| $CO_3(2-)$ | 0.00039017 | 0.00011673 | 0.00020317 |
| $OH(-)$ | 1.87E-07 | 1.49E-07 | 1.60E-07 |
| $NH_3$(aq) | 7.79E-07 | 1.36E-05 | 1.26E-05 |
| $NH_4(+)$ | 0.00039165 | 0.02585194 | 0.0256976 |
| $N_2$(aq) | 0.00273379 | 0.00317021 | 0.00263621 |
| $O_2$(aq) | 2.03E-09 | 1.31E-14 | 0 |
| $CH_4$(aq) | 0.00013211 | 3.01E-07 | 4.21E-07 |
| CO(aq) | 8.66E-06 | 1.24E-12 | 8.62E-14 |
| $H_2S$(aq) | 3.13E-05 | 0.00048445 | 0.00046404 |
| $SO_4(2-)$ | 0.0015666 | 0.00072908 | 0.00068588 |
| $H_2$(aq) | 6.48E-07 | 3.28E-11 | 4.19E-10 |
| **Available Gibbs energy, $\Phi$ (J/mol)** |  | **234** | **232** |

**table S5. Archean minimum disequilibrium.** Mixing ratios are reported for gaseous species, and moles per mole of atmosphere for aqueous species. Note that final Aspen values that do not completely conserve mass and charge are an artifact of Aspen automatically adjusting initial abundances slightly before performing Gibbs energy calculations.

|  | initial | final matlab | final - ASPEN |
|---|---|---|---|
| $H_2O$(liquid) | 436.77573 | 436.778199 | 437.0025 |
| $O_2$ | 1.00E-15 | 5.17E-13 | 0 |
| $N_2$ | 0.98 | 0.979146 | 0.9801331 |
| $H_2O$ | 2.00E-02 | 0.01647763 | 0.0166226 |
| $CO_2$ | 0.001 | 0.00042043 | 0.00057407 |
| $NH_3$ | 1.00E-15 | 1.80E-08 | 1.51E-08 |
| $CH_4$ | 0.0001 | 9.40E-09 | 7.91E-09 |
| $CO$ | 1.00E-15 | 8.58E-12 | 0 |
| $H_2$ | 1.00E-15 | 5.03E-08 | 4.85E-08 |
| $H_2S$ | 7.83E-15 | 7.83E-15 | 0 |
| $H(+)$ | 7.29E-08 | 9.70E-08 | 4.38E-08 |
| $Na(+)$ | 4.30721038 | 4.30721038 | 4.30721 |
| $Cl(-)$ | 4.28E+00 | 4.27587006 | 4.27587 |
| $HCO_3(-)$ | 0.02816377 | 0.02960221 | 0.0292059 |
| $CO_2$(aq) | 0.00031353 | 0.00015015 | 0.00019063 |
| $CO_3(2-)$ | 0.00158412 | 9.90E-04 | 0.00119208 |
| $OH(-)$ | 8.38E-06 | 1.47E-05 | 1.10E-05 |
| $NH_3$(aq) | 7.83E-15 | 1.37E-05 | 9.33E-06 |
| $NH_4(+)$ | 7.83E-15 | 0.00025639 | 0.00026079 |
| $N_2$(aq) | 0.00535822 | 6.08E-03 | 0.00509003 |
| $O_2$(aq) | 1.00E-15 | 5.01E-13 | 0 |
| $CH_4$(aq) | 1.32E-06 | 9.88E-11 | 9.18E-11 |
| $CO$(aq) | 1.00E-15 | 5.73E-12 | 0 |
| $H_2S$(aq) | 7.83E-15 | 7.83E-15 | 0 |
| $SO_4(2-)$ | 7.83E-15 | 7.83E-15 | 0 |
| $H_2$(aq) | 1.00E-15 | 2.21E-10 | 2.77E-10 |
| **Available Gibbs energy, $\Phi$ (J/mol)** |  | **5.1** | **3.4** |

**table S6. Reactions contributing to Proterozoic disequilibrium.** Note that the contributing Gibbs energies do not sum exactly to the total available energy because there are additional contributions from water activity changes, and because reactions are not independent.

| Description | Reaction | Contribution to maximum disequilibrium | Contribution to minimum disequilibrium |
|---|---|---|---|
| Nitrate formation | $5O_2 + 2N_2 + 2H_2O \rightarrow 4H^+ + 4NO_3^-$ | 640 J/mol (72%) | 7.4 J/mol (79%) |
| Carbonate speciation from ocean acidification | $HCO_3^- + H^+ \rightarrow H_2O + CO_2$ | | |
| Ammonium oxidation | $3O_2 + 4NH_4^+ \rightarrow 6H_2O + 4H^+ + 2N_2$ | 101 J/mol (11%) | 0 J/mol (0%) |
| Methane oxidation | $CH_4 + 2O_2 \rightarrow 2H_2O + CO_2$ | 75 J/mol (8%) | 2.1 J/mol (23%) |
| Sulfide oxidation | $H_2S + 2O_2 \rightarrow 2H^+ + SO_4^{2-}$ | 27 J/mol (3%) | 0 J/mol (0%) |
| Total available Gibbs energy | | 884 J/mol | 9.5 J/mol |

**table S7. Reactions contributing to Archean disequilibrium.** Note that the contributing Gibbs energies do not sum exactly to the total available energy because there are additional contributions from water activity changes, and because reactions are not independent. For instance, hydrogen oxidation produces methane that then adds to ammonium formation.

| Description | Reaction | Contribution to maximum disequilibrium | Contribution to minimum disequilibrium |
|---|---|---|---|
| Ammonium formation | $5CO_2 + 4N_2 + 3CH_4 + 14H_2O \rightarrow 8NH_4^+ + 8HCO_3^-$ | 170 J/mol (74%) | 4 J/mol (80%) |
| Carbon monoxide oxidation | $4CO + 2H_2O \rightarrow 3CO_2 + CH_4$ | 42 J/mol (18%) | 0 J/mol (0%) |
| Hydrogen oxidation | $CO_2 + 4H_2 \rightarrow 2H_2O + CH_4$ | 2 J/mol (1%) | 0 J/mol (0%) |
| Sulfate reduction | $CO_2 + CH_4 + SO_4^{2-} \rightarrow H_2S + 2HCO_3^-$ | 3 J/mol (1%) | 0 J/mol (0%) |
| Carbon speciation | $H_2O + CO_2 + CO_3^{2-} \rightarrow 2HCO_3^-$ | <1 J/mol (<1%) | 1 J/mol (20%) |
| Ammonia formation | $N_2 + 3H_2 \rightarrow 2NH_3$ | <1 J/mol (<1%) | 0 J/mol (0%) |
| Total available Gibbs energy | | 231 J/mol | 5.1 J/mol |

**table S8. Sensitivity of Archean disequilibrium to difficult-to-observe variables.**
Disequilibria values highlight in bold indicate scenarios where methane-depletion did not occur (<90% depletion). Available energies are reported for both our Matlab calculations, and the commercial software package Aspen Plus, and the two methods are in agreement.

|  |  | Available Gibbs free energy (J/mol) | |
|---|---|---|---|
|  |  | Matlab | ASPEN |
| Temperature, K | 273.15 | 369 | 359 |
|  | 288.15 | 234 | 232 |
|  | 298.15 | 220 | 204 |
| Ocean alkalinity, mmol/kg (pH) | 4 (pH=5.4) | 363 | 359 |
|  | 40 (pH=6.4) | 234 | 232 |
|  | 200 (pH=7.1) | 152 | **181** |
| Ocean salinity, relative to modern | 0.1 | 199 | 198 |
|  | 1 | 234 | 232 |
|  | 10 | 249 | 223 |
| Ocean volume, relative to modern | 0.1 | **89** | **89** |
|  | 0.5 | 188 | 187 |
|  | 1 | 234 | 232 |
|  | 2 | 279 | 277 |
|  | 10 | 405 | 401 |
|  | 50 | 701 | 682 |
| Atmospheric pressure, atm | 0.1 | 125 | 141 |
|  | 0.5 | 220 | 213 |
|  | 1 | 234 | 232 |
|  | 2 | 271 | 262 |
|  | 10 | 366 | 354 |
| Low pressure (0.5 bar) and variable bulk abundances | 17% $N_2$, 80% $CO_2$ | 198 | 191 |
|  | 50% $N_2$, 47% $CO_2$ | 220 | 213 |
|  | 77% $N_2$, 20% $CO_2$ | 224 | 214 |
| Low N2 abundance | 2% $N_2$, 95% $CO_2$ | 151 | 150 |
|  | 50% $N_2$, 47% $CO_2$ | 234 | 232 |